\documentclass[twocolumn]{aastex6}

\usepackage{ulem}

\slugcomment{Draft v7.0}

\shorttitle{Dimming of KIC 8462852}
\shortauthors{Meng et al. 2016}

\begin{document}


\title{Extinction and the Dimming of KIC 8462852}

\author{Huan Y. A. Meng\altaffilmark{1}, George Rieke\altaffilmark{1},   Franky Dubois\altaffilmark{2}, Grant Kennedy\altaffilmark{3}, Massimo Marengo\altaffilmark{4}, Michael Siegel\altaffilmark{5},  Kate Su\altaffilmark{1}, Nicolas Trueba\altaffilmark{4}, Mark Wyatt\altaffilmark{3}, Tabetha Boyajian\altaffilmark{6},  C. M. Lisse\altaffilmark{7}, Ludwig Logie\altaffilmark{2}, Steve Rau\altaffilmark{2}, Sigfried Vanaverbeke\altaffilmark{2,8}}

\altaffiltext{1}{Steward Observatory, Department of Astronomy, University of Arizona, 933 North Cherry Avenue, Tucson, AZ 85721}

\altaffiltext{2}{Astrolab IRIS, Verbrandemolenstraat, Ypres, Belgium and Vereniging voor Sterrenkunde, Werkgroep Veranderlijke Sterren, Belgium}

\altaffiltext{3}{Institute of Astronomy, University of Cambridge, Madingley Road, Cambridge CB3 0HA, UK}

\altaffiltext{4}{Department of Physics and Astronomy, Iowa State University, A313E Zaffarano, Ames, IA 50010}

\altaffiltext{5}{Department of Astronomy and Astrophysics, Pennsylvania State University, 525 Davey Lab, University Park, PA 16802}

\altaffiltext{6}{Department of Physics and Astronomy, Louisiana State University, 261-A Nicholson Hall, Tower Dr, Baton Rouge, LA 70803}

\altaffiltext{7}{Applied Physics Laboratory, Johns Hopkins University, 11100 Johns Hopkins Road, Laurel, MD 20723}

\altaffiltext{8}{Center for Mathematical Plasma Astrophysics, University of Leuven, Belgium}

\email{hyameng@lpl.arizona.edu}


\begin{abstract}
To test alternative hypotheses for the behavior of KIC 8462852, we obtained measurements of the star over a wide wavelength range from the UV to the mid-infrared from October 2015 through December 2016, using {\it Swift}, {\it Spitzer} and at AstroLAB IRIS. The star faded in a manner similar to the long-term fading seen in {\it Kepler} data about 1400 days previously. The  dimming rate for the entire period reported is $22.1 \pm 9.7$ milli-mag yr$^{-1}$ in the {\it Swift} wavebands, with amounts of $21.0 \pm 4.5$ mmag in the groundbased $B$ measurements, $14.0 \pm 4.5$ mmag in $V$, and $13.0 \pm 4.5$ in $R$, and a rate of $5.0 \pm 1.2$ mmag yr$^{-1}$ averaged over the two warm {\it Spitzer} bands. Although the dimming is small, it is seen at $\gtrsim$ 3 $\sigma$ by three different observatories operating from the UV to the IR.  The presence of long-term secular dimming means that previous SED models of the star based on photometric measurements taken years apart may not be accurate. We find that stellar models with $T_{eff} =$ 7000 - 7100 K and $A_V \sim 0.73$ best fit the {\it Swift} data from UV to optical. These models also show no excess in the near-simultaneous {\it Spitzer} photometry at 3.6 and 4.5 $\mu$m, although a longer wavelength excess from a substantial debris disk is still possible (e.g., as around Fomalhaut). The wavelength dependence of the fading favors a relatively neutral color (i.e., $R_V \gtrsim$ 5, but not flat across all the bands) compared with the extinction law for the general ISM ($R_V =$ 3.1), suggesting that the dimming arises from circumstellar material.

\end{abstract}



\section{Introduction}

KIC 8462852, also known as Boyajian's Star, is an enigmatic object discovered by citizen scientists of the Planet Hunters project studying data from the {\it Kepler} mission \citep{boyajian2016}. The main-sequence F1/2 V star \citep{lisse2015, boyajian2016} at $\sim$400 pc \citep{boyajian2016,gaia2017} has undergone irregularly shaped dips in flux up to $\sim$ 20\% with durations of one to a few days \citep{boyajian2016}. A new episode of dips has started in May- June, 2017 \citep{boyajian2017}. The star also faded throughout the  {\it Kepler} mission \citep{borucki2010}, initially in a slow decline, and 
then a more rapid fading by $\sim$ 2\% over about 300 days \citep{montet2016}.  Such behavior is virtually unique among normal main-sequence stars \citep{schlecker2016}. Archival data have also been used to suggest a  decline in stellar brightness over the past century with an average rate of $-0.151 \pm 0.012$\% yr$^{-1}$ \citep{schaefer2016}, though the existence and significance
 of the century-long trend are disputed \citep{hippke2016a,hippke2017,lund2016}. In addition, adaptive 
optics-corrected images in the $JHK$ bands reveal a nearby source 2\arcsec\ from the primary star, with 
brightness and color consistent with a M2 V companion at a projected distance of $\sim$800 AU \citep{boyajian2016}.

Peculiar light curves and slow trends are common among young stellar objects (YSOs) \citep{rebull2014,stauffer2015,ansdell2016}, and may result from the obscuration by dust generated by disintegrating planets \citep[e.g.,][]{rappaport2012,rappaport2014,sanchis-ojeda2015} or planetesimal/planet collisions \citep[e.g.,][]{meng2014,meng2015} if viewed edge-on \citep{bozhinova2016}. However, KIC 8462852 does not appear to fit into either scenario.
The optical to mid-IR spectrum of the star confirms that it is a mature main-sequence dwarf; spectral energy distribution (SED) modeling finds no significant IR excess in the 3-5 $\micron$ region that could arise from a warm debris disk \citep{marengo2015}; no excess is seen in the WISE photometry \citep{boyajian2016}; and  millimeter and sub-millimeter continuum observations also find no significant excess emission towards the star \citep{thompson2016}. 

To characterize the effects of apparent astrometric motions that are driven by variability of field stars, the {\it Kepler} data have been studied using a principal component technique to remove correlated trends that are not relevant to the phenomena under investigation \citep{makarov2016}.  It was found that some variations seen in the {\it Kepler} light curve are likely from other sources close to the line of sight of KIC 8462852. In particular, the study suggests that the 0.88-day period, presumed to be the rotational modulation of KIC 8462852 \citep{boyajian2016}, is likely from a contaminating source. While the major dips in Q8 and Q16/17 (and the long-term secular dimming) are confirmed to be from KIC 8462852, the origin of the smaller dips is less certain.

A viable explanation for the bizarre dips in the light curve of KIC 8462852 is the apparition of a large family of comets \citep{boyajian2016}, possibly the onset of a period like the Late Heavy Bombardment \citep{lisse2015}. \citet{bodman2016} show that the hypothesis is plausible by successfully modeling the last episode of dimming events in {\it Kepler} Quarters 16 and 17. Similar simulations by \citet{neslusan2017} also reproduce the primary features of the dips with one dust-enshrouded planetary object for each dip. However, this latter type of model would require additional comets or dust-enshrouded planets to explain any additional dips, essentially adding a large number of new free parameters. 

A number of other possibilities have been proposed.  A possible explanation is that one or more planetary bodies have spiraled into the star. It is speculated that the return of the star to thermal equilibrium may explain the slow dimming, while the deep dips may arise from transits of planetary debris \citep{metzger2017}. \citet{ballesteros2017} model the events as being due to a  Trojan-like asteroid system orbiting the star. In contrast, \citet{wright2016} suggest that the dimming might be caused by foreground dust in the ISM, with dense clumps in an intervening dark cloud responsible for the deep dips. Extinction by clumpy material in the outer Solar System has also been suggested \citep{katz2017}. In addition, instabilities in the star itself have been proposed \citep{sheikh2016,foukal2017}.

To investigate this mystery, we are conducting ongoing monitoring of KIC 8462852 and its surrounding field with two space telescopes in seven wavebands, {\it Swift}/UVOT in $uvw2$ (effective wavelength 2030 \AA), $uvm2$ (2231 \AA), $uvw1$ (2634 \AA), $u$ (3501 \AA), and $v$ (5402 \AA) bands \citep{poole2008}, and {\it Spitzer}/IRAC at 3.6 and 4.5 $\micron$ \citep{fazio2004}. In this paper, we report the results from the monitoring from October 2015 through December 2016 (we include a few measurements past this cutoff but have not made use of them in the analysis). We also use the automated and homogeneous observations from the AAVSO database obtained through December 2016 in optical $BVR$ bands with the Keller F4.1 Newtonian New Multi-Purpose Telescope (NMPT) of the public observatory AstroLAB IRIS, Zillebeke, Belgium.

\section{Observations}

In this section, we describe the basic observations obtained with {\it Swift}/UVOT, AstroLAB IRIS, and {\it Spitzer}. Each of these sets of data indicates a subtle dimming of the star. However, gaining confidence in this result requires a detailed analysis of calibration issues, which is reserved for Section 3. 

\subsection{{\it Swift}/UVOT}

The Ultraviolet/Optical Telescope (UVOT) is one of three instruments aboard the {\it Swift} mission. It is a modified Ritchey-Chr\'etien 30 cm telescope with a wide ($17\arcmin \times 17\arcmin$) field of view and a microchannel plate intensified CCD detector operating in photon-counting mode \citep[see details in][]{roming2000,roming2004,roming2005}. UVOT provides images at 2.3\arcsec\ resolution and includes a clear white filter, $u$, $b$, and $v$ optical filters, $uvw1$, $uvm2$, and $uvw2$ UV filters, a magnifier, two grisms, and a blocking filter. The $uvw2$ and $uvw1$ filters have substantial red leaks, which have been characterized to high precision by \citet{breeveld2010} and are included in the current UVOT filter curves. Calibration of the UVOT is discussed in depth by \citet{poole2008} and \citet{breeveld2011}.

In full-frame mode, the CCD is read every 11 ms, which creates a problem of coincidence loss (similar to pile-up in the X-ray) for stars with count rates greater than 10 cts s$^{-1}$ ($\sim$15 mag, depending on filter).  The camera can be used in a windowed mode, in which a subset of the pixels is read. Given the brightness of KIC 8462852, to reduce the coincidence corrections we observed in a $5\arcmin \times 5\arcmin$ window (70 $\times$ 70 pixels), resulting in a 3.6 ms readout time. The observations were generally 1 ks in duration, utilizing a mode that acquired data in five filters ($v$, $u$, $uvw1$, $uvm2$ and $uvw2$, from 5900 to 1600 \AA). To improve the precision of the photometric measurements and ensure that the target star and comparison stars all landed in the readout window, observations were performed with a ``slew in place,'' in which {\it Swift} observed the field briefly in the ``filter of the day" -- one of the four UV filters -- before slewing a second time for more precise positioning. The slew-in-place images were used for additional data points in the UV. KIC 8462852 was first observed on October 22, 2015 and then approximately every three days from December 4, 2015 to March 27, 2016. It was later observed in coordination with the {\it Spitzer} campaign starting August, 2016.

X-ray data were obtained simultaneously with the {\it Swift}/XRT \citep[X-ray Telescope,][]{burrows2005}. 
No X-ray emission within the passband from 0.2 to 10 keV is seen from KIC 8462852 in 52 ks of exposure time down to a limit of $5 \times 10^{-15}$ erg s$^{-1}$ cm$^{-2}$ using the online analysis tools of \citet{evans2009}.

\subsection{AstroLAB IRIS}

Optical observations in $B$, $V$, and $R$ bands were taken with the 684 mm aperture Keller F4.1 Newtonian New Multi-Purpose Telescope (NMPT) of the public observatory AstroLAB IRIS, Zillebeke, Belgium. The CCD detector assembly is a Santa Barbara Instrument Group (SBIG) STL 6303E operating at $-20{\degr}$C. A 4-inch Wynne corrector feeds the CCD at a final focal ratio of 4.39, providing a nominal field of view of $20\arcmin \times 30\arcmin$. The 9 $\micron$ physical pixels project to 0.62\arcsec\ and are read out binned to $3 \times 3$ pixels, i.e., 1.86\arcsec\ per combined pixel. The $B$, $V$, and $R$ filters are from Astrodon Photometrics, and have been shown to reproduce the Johnson/Cousins system closely \citep{henden2009}. The earliest observation was made on September 29, 2015. There is a gap in time coverage from January 8 to June 8, 2016. We report the observations through December 2016.

\subsection{{\it Spitzer}}

{\it Spitzer} observations are made with the Infrared Array Camera (IRAC) with a uniform exposure design, which uses a cycling dither at 10 positions on the full array with 12 s frame time. Using multiple dither positions tends to even out the intra- and inter-pixel response variations of the detector; repeating the same dither pattern at every epoch puts KIC 8462852 roughly on the same pixels of the detector, further reducing potential instrumental bias on the photometry\footnote{\hspace{-10mm}\url{http://ssc.spitzer.caltech.edu/warmmission/news/}
\url{18jul2013memo.pdf}}. 
Our first {\it Spitzer} observation was executed on January 16, 2016. There is a gap in the time baseline of the monitoring in the period from April to July 2016 (from MJD 57475 to 57605) when KIC 8462852 was out of the visibility window of {\it Spitzer}\footnote{Note that the time coverage gap in the {\it Swift} and {\it Spitzer} data is only partially overlapped with the gap in the ground-based data.}. The {\it Swift}-{\it Spitzer} coordinated monitoring is still underway at the time this paper was written. 

\begin{deluxetable*}{cccccccc}
\tabletypesize{\scriptsize}
\tablecaption{UVOT Photometry of KIC 8462852
\label{photom}}
\tablehead{
\colhead{Waveband}& \colhead{MJD\tablenotemark{a}} & \colhead{raw} & \colhead{$\sigma$} & \colhead{corrected}  & \colhead{$\sigma$} & \colhead{relative comparison\tablenotemark{b}} & \colhead{$\sigma$}  \\
\colhead{}& \colhead{} & \colhead{magnitude} & \colhead{magnitude} & \colhead{magnitude}  & \colhead{magnitude} & \colhead{magnitude} & \colhead{magnitude}  
}
\startdata
$uvw2$	&	57317.518	&	14.820	&	0.050	&	14.776	&	0.089	&	0.045	&	0.074	\\
$uvw2$	&	57317.521	&	14.800	&	0.030	&	14.793	&	0.055	&	0.007	&	0.046	\\
$uvw2$	&	57317.552	&	14.790	&	0.030	&	14.787	&	0.047	&	0.003	&	0.036	\\
$uvw2$	&	57317.554	&	14.810	&	0.030	&	14.784	&	0.043	&	0.026	&	0.030	\\
$uvw2$	&	57317.584	&	14.780	&	0.030	&	14.755	&	0.058	&	0.025	&	0.050	\\
$uvw2$	&	57317.587	&	14.820	&	0.030	&	14.735	&	0.061	&	0.085	&	0.053	\\
\enddata
\tablenotetext{a}{Modified Julian date}
\tablenotetext{b}{Average brightening of the comparison star measurements}
\tablecomments{This table is available in its entirety in a machine-readable form in the online journal and as an appendix to this posting. A portion is shown here for guidance regarding its form and content.}
\end{deluxetable*}

\section{Data Processing and Analysis }

Our three data sources provide multiple accurate measurements of KIC 8462852 over a year. They were all interrupted when the viewing angle to KIC 8462852 passed too close to the sun. Because of differing viewing constraints, exactly when this gap in the data occurs differs among the observatories. Because of the differences in time coverage, 
we analyze long-term trends in the data sets in two ways. First, within a given data set, we fit a linear trend and use the slope and its error as an indication of any change.  The {\it Swift} measurements are mostly prior to the gap, so we fitted both before the gap and for the whole set of measurements.  For the sake of comparison, we treat the groundbased data the same way. The {\it Spitzer} observations began at the end of the first groundbased sequence, so we only fit the whole set. 

Although these fits make use of all the data in each band, they may give misleading information on the color behavior because the data do not have identical time coverage. In discussing color trends we focus just on the UVOT and groundbased data obtained in overlapping time sequences. For the {\it Spitzer} data, we calculate the difference from the first measurement to the later ones, and compare with a similar calculation for the groundbased point closest in time to the first {\it Spitzer} point, relative to the post-gap results from the ground. Details of these procedures are given below.  

\subsection {{\it Swift}  Data\label{trend}}

{\it Swift}/UVOT data were obtained directly from the HEASARC archive\footnote{\url{http://heasarc.gsfc.nasa.gov/cgi-bin/W3Browse/swift.pl/}. The same processed UVOT data are also available through the MAST archive at STScI, \url{http://archive.stsci.edu}.}. We then used the HEASARC FTOOLS software\footnote{\url{http://heasarc.gsfc.nasa.gov/docs/software/lheasoft/}}
program UVOTSOURCE on the transformed sky images to generate point source photometry. 

The Swift/UVOT photometry includes a filter-dependent correction that accounts for the decline in instrument sensitivity \citep{breeveld2011}.  However, the potential fading of KIC 8462852 pushes the boundaries of the UVOT calibration, which is specified to within 1\%.  To check for any residual sensitivity changes, we made photometric measurements for four additional field stars in the UVOT field. The reference stars selected are KIC 8462934, KIC 8462763, KIC 8462843, and KIC 8462736, which all lie 1\farcm4 to 2\farcm0 from KIC 8462852. Since these reference stars are all fainter than KIC 8462852 by 2 $-$ 5 magnitude in the UV and have larger photometric errors individually, we took their weighted average as our photometric reference to minimize the noise. Their linear fits indicate that the average reference star appears to get brighter in the UVOT data over the full range of our time coverage, especially when the post-gap data are considered (-22.2 $\pm$ 5.2 mmag yr$^{-1}$). We examined the reference stars individually and found that all four of them follow similar and consistent brightening rates, eliminating the possibility of "bad" star contamination. This suggests a small residual instrumental trend in the UVOT data, which could reflect a small overestimate of the sensitivity loss or a small residual in the coincidence loss correction. 

\begin{figure*}
\epsscale{0.7}
\plotone{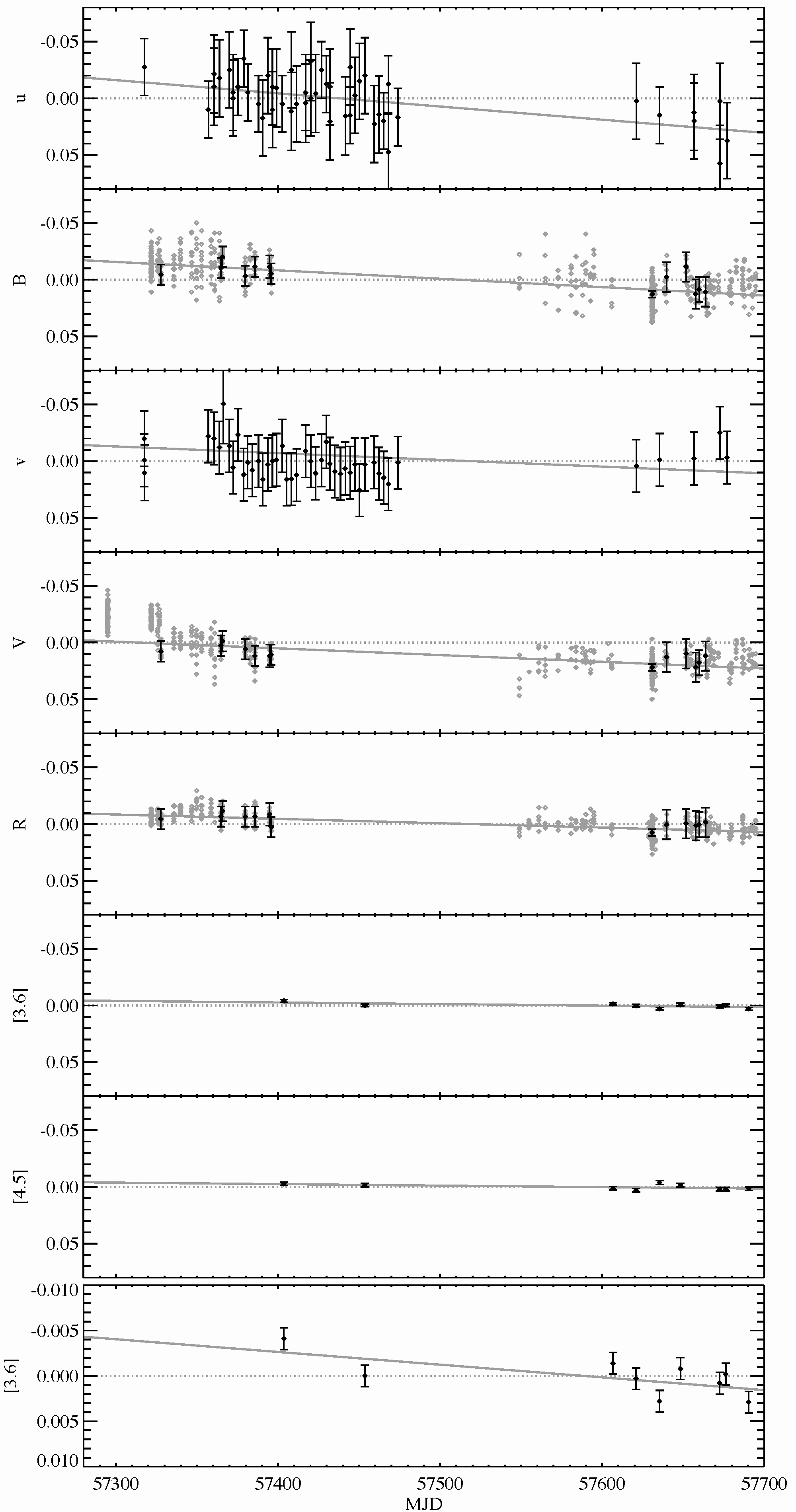}
\caption{Demonstration of the fading trend of KIC 8462852 in {\it Swift}/UVOT $u$ and $v$ bands, ground-based $BVR$ bands, and {\it Spitzer}/IRAC 3.6 and 4.5 $\micron$, sorted by increasing wavelength in panels 1 - 7 from top to bottom. Panel 8 repeats the results at [3.6] on an expanded scale. The data sequences extend from October, 2015 through December, 2016. For the $B$, $V$, and $R$ bands, the light gray points without error bars are the individual measurements after rejecting only nights with indicated large errors in the photometry of KIC 8462852. The photometry in all wavebands is normalized to the average magnitude in the band and shown to the same vertical axis scale. For illustrative purposes only, in each panel the solid gray line is a linear fit to indicate the dimming; a level line is shown in dotted gray for comparison.The dark points with error bars for $B$, $V$, and $R$ are the averages for the nights showing a high degree of uniformity in the reference star measurements and are the most reliable.  
\label{lightcurve}}
\end{figure*}

To remove the instrumental trend of {\it Swift}/UVOT, we subtract the normalized magnitude of the average reference star from the absolutely calibrated magnitude of KIC 8462852. This inevitably propagates the photometric uncertainties of the average reference star into the KIC 8462852 light curve.  In all following discussions, we only use the corrected calibration data of {\it Swift}/UVOT. All the {\it Swift} photometry is given in Table~\ref{photom} and is displayed in Figure~\ref{lightcurve}.
 For a first search for trends in the brightness of the star, we performed a linear fit to the photometry. As shown in Table~2,  
with the corrected calibration the fading of KIC 8462852 is seen at a 3 $\sigma$ significance level in the pre-gap data and by $>$ 2 $\sigma$ in the full data set.

\begin{deluxetable*}{ccccccccccc}
\tabletypesize{\footnotesize}
\tablecaption{Trends of the brightness of KIC 8462852}
\label{trends}
\tablehead{
 & & \multicolumn{4}{c}{Pre-Gap} & & \multicolumn{4}{c}{Full Data} \\
\cline{3-6} \cline{8-11}
\colhead{Waveband} & \colhead{$\lambda_{eff}$} & \colhead{$dm/dt$} & \colhead{$\sigma$($dm/dt$)} & \colhead{$\chi_{red}^2$} & \colhead{sig\tablenotemark{a}} & & \colhead{$dm/dt$} & \colhead{$\sigma$($dm/dt$)} & \colhead{$\chi_{red}^2$} & \colhead{sig\tablenotemark{a}} \\
\colhead{} & \colhead{($\micron$)} & \colhead{(mmag yr$^{-1}$)} & \colhead{(mmag yr$^{-1}$)} & & & & \colhead{(mmag yr$^{-1}$)} & \colhead{(mmag yr$^{-1}$)} & & 
}
\startdata
$uvw2$ & 0.2030 & 83.7 & 66.6 & 0.81 & & & 71.8 & 31.5 & 0.92 \\
$uvm2$ & 0.2231 & 210.5 & 203.9  & 1.03 & & & -60.3 & 86.5 & 1.07 \\
$uvw1$ & 0.2634 & 108.5 & 61.5 & 0.96 & & & 17.4 & 24.5 & 1.04 \\
$u$ & 0.3501 & 69.0 & 44.0 & 0.29 & & & 42.7 & 17.8 & 0.30 \\
$v$ & 0.5402 & 54.9 & 32.7 & 0.27 & & & 1.9 & 14.5 & 0.34 \\
\multicolumn{2}{c}{{\it Swift} Average\tablenotemark{b}} & 71.0 & 22.6 & \nodata & Y & & 22.1 & 9.7 & \nodata & \\
$B$ & 0.435 & 20.2 & 11.4 & \nodata & & & 26.3 & 1.5 & \nodata & Y \\
$V$ & 0.548 & 16.2 & 8.1  & \nodata &  & & 21.6 & 1.5\tablenotemark{d} &\nodata & Y\\
$R$ & 0.635 & 1.6 & 6.7 & \nodata & & & 13.1 & 1.0 & \nodata & Y \\
$[3.6]\tablenotemark{c}$ & 3.550 & \nodata  &\nodata  & \nodata & & & 5.1 & 1.5 & 1.99 & Y \\
$[4.5]\tablenotemark{c}$ & 4.493 & \nodata &\nodata & \nodata & & & 4.8 & 2.0 & 1.84 & \\
\multicolumn{2}{c}{{\it Spitzer} Average\tablenotemark{b}} & \nodata & \nodata & \nodata & & & 5.0 & 1.2 & \nodata & Y \\
\enddata
\tablenotetext{a}{Significance flag. ``Y'' means that the magnitude changing rate is significantly different from zero at the  3$\sigma$ level or higher. Otherwise this is left blank.}
\tablenotetext{b}{Averaging the data ignoring the wavelength information implicitly assumes a grey color of the fading.}
\tablenotetext{c}{We only have 2 epochs of {\it Spitzer} observations before the gap, and the gap is offset relative to the groundbased data, so we do not quote values pre-gap.}
\tablenotetext{d} {Computed omitting the first night, which is systematically high and had a large number of measurements that drive the slope inappropriately.}

\end{deluxetable*}

\subsection{Ground-based Data}

The {\it Swift} data suggest that the star faded during our observations, and such behavior would be consistent with the long-term secular fading of KIC 8462852 observed previously \citep{schaefer2016,montet2016}. To probe this behavior further, we turn to the ground-based  BVR photometry obtained at AstroLAB IRIS and available from AAVSO. We did not detect any major dips in the flux from the monitoring up to December 2016, but our measurements do indicate a slight long-term secular dimming. 

We used differential photometry relative to four stars in the field, selected to be similar in brightness and color to KIC 8462852  (see  Table~\ref{refstars}). These observations were obtained simultaneously with those of KIC 8462852, typically in a series in $V$ followed by $B$ and then $R$. The reductions utilized the LesvePhotometry reduction package \citep{deponthiere2013}, which is optimized for time-series observations of variable stars. It automates reduction of the data in a series of observations, providing a homogeneous database of photometry for KIC 8462852.  Errors from photon noise on the source, scintillation noise and background noise are determined for each observation, using the methodology in \citet{newberry1991}. We eliminated data from a single night when the source was at large ($>$ 2) airmass. We also eliminated data from nights with larger than normal estimated errors ($>$ 0.025 magnitudes rms) for KIC 8462852, as identified either by the reduction package or by large rms errors for the data obtained within that night. The remaining photometry in $B$, $V$, and $R$ bands is displayed in Figure~\ref{lightcurve}. The scatter is somewhat larger than implied by the internal error estimates; therefore, we will base our error estimates on the scatter. Linear fits show a dimming in all three colors over the full data set, and probable dimming but not at signficant levels pre-gap (see Table~2). No significant dimming is seen post-gap.  In estimating the uncertainties of these slopes, we found that the reduced $\chi^2$ using the reported internal errors was large (2 to 4 depending on the band), consistent with our finding that the rms scatter of the measurements is larger than the reported errors. We brought the reduced $\chi^2$ of the fit to $\sim$ 1 by adding an additional error of 0.01 mag in quadrature to each measurement. 

\begin{deluxetable}{cccccccccc}
\tablecaption{Comparison stars for BVR photometry
\label{refstars}}
\tablehead{
\colhead{ID}& \colhead{RA  (2000) } & \colhead{DEC (2000)} & \colhead{V} & \colhead{B-V}  & \colhead{V-R} 
}
\startdata
KIC 8462852 & 20 06 15.46 & +44 27 24.6  & 11.86 & 0.54 & 0.40 \\
Star 1 & 20 07 09.07 & +44 20 17.1 & 11.59 & 0.54 & 0.40 \\
Star 2 & 20 06 01.24 & +44 29 32.4 & 12.42 & 0.79 & 0.51  \\
Star 3 & 20 06 21.21 & +44 30 52.2 & 12.81 & 0.51 & 0.40  \\
Star 4 & 20 06 48.09 & +44 22 48.1 & 11.26 & 0.47 & 0.35  \\
\enddata
\end{deluxetable}

There are two issues with the linear fits. The first is that, given that there is no evidence for fading in the data after the interruption due to solar viewing constraints (see  Figure~\ref{lightcurve}), the fits tend to be high toward the beginning and low toward the end of the post-gap sequence. This is particularly prominent in the $v$ and $V$ fits as shown in  Figure~\ref{lightcurve}. The linear fits are meant as the simplest way to quantify the dimming that would include all the data in each band, but they appear not to be exactly the correct dependence.

The second issue is that, for the groundbased data, our discussion above does not include the possibility of systematic errors affecting the comparison pre- and post-gap, effects that are not included in the  LesvePhotometry package. (We have already eliminated such errors for the $\it Swift$/UVOT data and, as discussed below, they should be negligible for the {\it Spitzer} measurements.) To test for such effects, we turned to the photometry of the four reference stars to estimate night-to-night and longer-term  errors and to select the nights with the most consistent observations to see if the dimming was apparent just using this subset of best measurements. 
Since these further steps made no reference to the photometry of KIC 8462852, they should introduce no bias in its measurements. 

We first computed the standard deviations of running sets of 45 measurements for each of these four stars. When this value exceeded an average of 0.015 per star, we investigated the photometry involved and eliminated nights contributing disproportionately to the value. Following this step, we examined the consistency of the remaining measurements of the reference stars. Since we do not know the ``true'' magnitudes of the stars, we instead tested for the consistency of the measurements of each star across the gap in time coverage. To establish a baseline, we identified two long consecutive sets of measurements for each star and each color, one on each side of the gap, that agreed well. We then tested each of the additional nights of data to see if they were consistent with this baseline, and added in the data for the nights that did not degrade the agreement across the gap. We carried out this procedure individually for each of the three colors, but found that the same nights were identified as having the highest quality photometry in each case. The final typical mis-match in photometry across the gap was 0.0032 magnitudes, showing that this vetting was effective in identifying nights with consistent results for the four reference stars. The photometry of KIC 8462852 on these nights is listed in Table~\ref{BVR} and plotted in Figure ~\ref{lightcurve}. 

The photometry selected to be of highest internal consistency is generally consistent with the rest of the measurements. The final averages for KIC 8462852 just based on these nights before and after the gap in the time series are shown in  Table~\ref{AAVSO}. Each band shows a small but statistically significant dimming from pre-gap to post-gap, both in the initial photometry (Table~2) and in that selected to be of highest quality.   

There is a hint of fading in the pre-gap data, but averaged over the three bands the net change is $0.010 \pm 0.005$ magnitudes, i.e. small and potentially insigificant. The post-gap data indicate no significant long-term secular changes beyond the errors of $\sim$ 0.003 magnitudes in any of the bands. These results suggest that most of the change in brightness occurred while the star was in the gap for the groundbased photometry. This behavior would be consistent with that observed for long-term dimming using {\it Kepler} data, where most of the change is a drop in brightness by $\sim$ 2\% over a period of 300 days \citep{montet2016}. The gap in our data is intriguingly about 1400 days ($\sim$ twice the interval between the two large dips in the light curve) past the time of the similar dimming seen in the Kepler data. The amplitude in the $V$ and $R$ bands (which together approximate the {\it Kepler} spectral response) is about $1.4 \pm 0.3$\% (Table 5), and the duration of the gap in our highest quality BVR data is $\sim$ 150 days, values that are also reminiscent of the Kepler observations.

\begin{deluxetable*}{ccccccc}
\tablecaption{Selected High-Quality BVR photometry
\label{BVR}}
\tablehead{
\colhead{JD} & \colhead{B (mag)} & \colhead{err\tablenotemark{a}} & \colhead{V (mag)} & \colhead{err\tablenotemark{a}}  & \colhead{R(mag)} & \colhead{err\tablenotemark{a}} 
}
\startdata
2457328.25	&  12.379 &	0.009 & 11.846 & 0.009 & 11.455 & 0.009  \\
2457365.29	&  12.373 &  0.009 & 11.841 & 0.009 & 11.453 & 0.009  \\
2457366.23	 &  12.363 & 0.009 & 11.837 & 0.009 & 11.448 & 0.009  \\ 
2457380.25	  &  12.380 & 0.009 & 11.844 & 0.009 & 11.453 & 0.009  \\
2457386.26	 &  12.372 & 0.009 & 11.850 & 0.009 & 11.453 & 0.009  \\
2457395.26	 &  12.372 & 0.010 & 11.850 & 0.010 & 11.451 & 0.010  \\
2457396.25	 &  12.378 & 0.009 & 11.849 & 0.009 & 11.462 & 0.009  \\
2457631.38	 &  12.396 & 0.003 & 11.860 & 0.003 & 11.467 & 0.003  \\
2457640.31	 &  12.381 & 0.013	& 11.851 & 0.013 & 11.460 & 0.013  \\
2457652.3	 &  12.372 & 0.013 & 11.848 & 0.013 & 11.459 & 0.013  \\
2457658.29	 &  12.396 & 0.013 & 11.860 & 0.013 & 11.461 & 0.013  \\
2457660.37	 &  12.392 & 0.011 & 11.856 & 0.011 & 11.460 & 0.011  \\
2457664.36	 &  12.394 & 0.013 & 11.850 & 0.013 & 11.458 & 0.013  \\
\enddata

\tablenotetext{a}{Combined rms errors of the mean, i.e. rms scatter divided by the square root of (n-1) where n is the number of measurements.}
\end{deluxetable*}

\begin{deluxetable*}{ccccccc}
\tablecaption{Summary of ground-based monitoring of KIC 8462852
\label{AAVSO}}
\tablehead{
\colhead{MJD Range}& \colhead{$\langle B\rangle$\tablenotemark{a}} & \colhead{$err(B)$\tablenotemark{b}} & \colhead{$\langle V\rangle$\tablenotemark{a}} & \colhead{$err(V)$\tablenotemark{b}}  & \colhead{$\langle R\rangle$\tablenotemark{a}} & \colhead{$err(R)$\tablenotemark{b}}
}
\startdata
57322 - 57396 & 12.374 & 0.0032 & 11.845 & 0.0032 & 11.453 & 0.0032 \\
57549 - 57693 & 12.395 & 0.0032 & 11.859 & 0.0032 & 11.466 & 0.0032 \\
Differences & 0.021 & 0.0045 & 0.014 & 0.0045 & 0.013 & 0.0045 \\
\enddata
\tablenotetext{a}{Average magnitude.}
\tablenotetext{b}{Combined rms errors of the mean, i.e. rms scatter divided by the square root of (n-1) where n is the number of measurements.}
\end{deluxetable*}

\subsection{{\it Spitzer}  Data}

For the purpose of probing for a long-term secular trend, we do not consider the earlier {\it Spitzer} photometry \citep{marengo2015}. That observation was made under the SpiKeS program (Program ID 10067, PI M. Werner) in January 2015, too far from the epochs of our new data. In addition, the SpiKeS observation was executed with an AOR design different from ours for the dedicated monitoring of KIC 8462852, which may lead to different instrumental systematics in the photometry.

The photometry we did use is from AORs 58782208, 58781696, 58781184, 58780928, 58780672, 58780416, and 58780160 (PID 11093, PI K. Y. L. Su) and 58564096 and 58564352 (PID 12124, PI Huan Meng). 
Photometric measurements were made on cBCD (artifact-corrected basic calibrated data) images with an aperture radius of 3 pixels and sky annulus inner and outer radii of 3 and 7 pixels, with the pixel phase effect and array location dependent response functions corrected. Aperture correction factors are 0.12856 and 0.12556 magnitudes at 3.6 and 4.5 $\micron$, respectively \citep{carey2012}. Individual measurements were averaged for each epoch. The photometry is summarized in Table~\ref{Spitzer}. 
During the period when the KIC 8462852 data were obtained, the photometric performance of IRAC is expected to have varied by 
less than 0.1\% per year (i.e., $<$ 1 mmag yr$^{-1}$)  \citep{irsa2015}.  There is a suggestion of a fading at the $\sim$ 3 $\sigma$ level of significance in the 3.6 $\mu$m band; there is also a fading in the 4.5 $\mu$m band at lower significance.

\begin{deluxetable}{ccccc}
\tablecaption{{\it Spitzer} photometry
\label{Spitzer}}
\tablehead{
\colhead{MJD} & \colhead{[3.6] (mag)} & \colhead{err} & \colhead{[4.5] (mag)} & \colhead{err}   
}
\startdata
57040.367  &   10.4627 &   0.0022 &  10.4243 &   0.0029  \\
57403.587  &   10.4510 &   0.0012 &   10.4334 &   0.0015  \\
57453.618  &   10.4551 &   0.0012 &   10.4346 &   0.0016  \\
57606.748  &   10.4537 &   0.0012 &   10.4375 &   0.0016  \\
57621.036  &   10.4554 &   0.0012 &   10.4392 &   0.0016  \\
57635.443  &   10.4579 &   0.0012 &   10.4324 &   0.0017  \\
57648.371  &   10.4543 &   0.0012 &   10.4345 &   0.0016  \\
57672.538  &   10.4559 &   0.0012 &   10.4382 &   0.0016  \\
57676.320  &   10.4549 &   0.0012 &   10.4383 &   0.0016  \\
57690.430  &   10.4580 &   0.0012 &   10.4377 &   0.0016  \\
57704.534  &   10.4574 &   0.0012 &   10.4363 &  0.0016  \\
57719.375  &   10.4604 &   0.0012 &   10.4357 &  0.0016  \\
57732.497  &   10.4514 &   0.0012 &   10.4392 &  0.0016  \\
57756.319  &   10.4486 &   0.0012 &   10.4344 &  0.0015  \\
57813.968  &   10.4511 &   0.0012 &    10.4388 &  0.0016  \\ 
\enddata
\end{deluxetable}

\subsection{Color-dependence of the dimming}

The three independent sets of observations presented in this paper all show evidence of dimming in KIC 8462852.  Measurements reported by the All-Sky Automated Survey for Supernovae (ASAS-SN) \citep{shappee2014, kochanek2017} do not completely overlap with ours, but a preliminary analysis shows that they show a fading of about 8 mmag at $V$, comparing their data between MJD of 57200 and 57334 with that between MJD of 57550 and 57740; their measurements may indicate a further small fading after that. The  ASAS-SN photometry has not been tested thoroughly for systematic errors at this small level (K. Stanek, private communication, and see warning at \url{https://asas-sn.osu.edu/}); nonetheless, the results agree within the mutual errors with ours. The dimming is also corroborated in measurements by \citet{gary2017}.

However, we find that the amount of dimming is significantly less 
in the infrared than in the optical and ultraviolet, as shown in  Figure~\ref{lightcurve}. We will investigate in 
Section 4 what constraints the wavelength dependence lets us put on this event. To do so, we need to focus on periods when measures are available in all the relevant bands (see  Figure~\ref{lightcurve}), since otherwise wavelength-independent and -dependent brightness changes are degenerate. 

The first sequence of Swift measurements extends well beyond the end date for the first sequence of BVR measurements. 
For the purposes of Section 4, we compute the dimming using the average of the measurements through MJD 57396. For the groundbased measurements, we take only the measurements on nights that passed our test for consistency of the comparison star measurements. We average all the data from such nights pre-gap and, separately, post-gap, and base the errors on the rms scatter of the individual measurements. For {\it Spitzer}, there is only a single measurement close in time to the first groundbased sequence, namely at JD 2457404.  To analyze these data, we first confirm the errors by computing the scatter in both bands (combined). This calculation indicates an error of 0.002, slightly larger than the quoted errors of 0.0012 - 0.0016. Using our more conservative error estimate, the change from the first point to the average of the post-gap ones at 3.6 $\mu$m is 0.0049 $\pm$ 0.0021 mag, and at 4.5 $\mu$m it is 0.0033 $\pm$ 0.0021 mag, or an average of 0.0041 $\pm$ 0.0015. We  compare this value with the change between the average of the two sets obtained from the ground closest in time to the first {\it Spitzer} one, namely at MJD 57395 and 57396 (both of which passed our tests for high quality data), versus the later (post-gap) groundbased measurements. We averaged the measurements in $B$ and $V$  together into a single higher-weight point for these two nights and then compared with the similar average of the post-gap measurements. The net change is 0.012 $\pm$ 0.0023 magnitudes, i.e., significantly larger than the change in the infrared.

 Table 7 summarizes the measurements we will use to examine the color-dependence of the dimming of KIC 8462852. It emphasizes conservative error estimation (including systematic ones) and homogeneous data across the three observatories, at the cost of nominal signal to noise. The dimming is apparent, at varying levels of statistical significance, in every band. The values in the table also agree with the slopes we computed previously (and noting that the time interval for the differences is about 74\% of a year), as shown in the table for the cases with relatively high weight values so comparisons are meaningful.

\begin{deluxetable*}{cccccccccc}
\tabletypesize{\scriptsize}
\tablecaption{Color dependence of the dimming
\label{color}}
\tablehead{
\colhead{Band}& \colhead{Wavelength ($\mu$m)} & \colhead{Interval} & \colhead{magnitude\tablenotemark{a}} & \colhead{Dimming}  & \colhead{Error\tablenotemark{b}} & \colhead{Telescope}
}
\startdata
uvw2 & 0.203 & pre-gap & 14.809 & --- & --- & Swift \\
uvw2 & 0.203 & post-gap & 14.826 & 0.017 & 0.018 & Swift \\
uvm2 & 0.223 & pre-gap & 14.808 & --- & --- & Swift \\
uvm2 & 0.223 & post-gap & 14.810 & 0.002 & 0.037 & Swift \\
uvw1 & 0.263 & pre-gap & 13.635 & --- & --- & Swift \\
uvw1 & 0.263 & post-gap & 13.649 & 0.014 & 0.011 & Swift \\
u & 0.346 & pre-gap & 12.575 & --- & --- & Swift \\
u & 0.346 & post-gap & 12.595 & 0.020 & 0.006 & Swift \\
u  &  0.346  & --- &  --- &  0.0316  &   0.0132  &  from slope calculation \\
B & 0.435 & pre-gap & 12.374 & --- & --- & AstroLAB-IRIS \\
B & 0.435 & post-gap & 12.395 & 0.021 & 0.0045 & AstroLAB-IRIS \\
B & 0.435 &  ---   &   ---  &   0.0195 &  0.0015  &  from slope calculation \\ 
v & 0.547 & pre-gap & 11.894 & --- & --- & Swift \\
v & 0.547 & post-gap & 11.904 & 0.010 & 0.006 & Swift \\
V & 0.548 & pre-gap & 11.845 & --- & --- & AstroLAB-IRIS \\
V & 0.548 & post-gap & 11.859 & 0.014 & 0.0045 & AstroLAB-IRIS \\
V & 0.548  &  ---   &   ---   &   0.0160  &  0.0015  &   from slope calculation \\
R & 0.635 & pre-gap & 11.453 & --- & --- & AstroLAB-IRIS \\
R & 0.635 & post-gap & 11.466 & 0.013 & 0.0045 & AstroLAB-IRIS \\
R & 0.635 &  ---  &   ---   &   0.0097  &  0.001  &  from slope calculation \\
{[3.6]} & 3.6 & pre-gap & 10.4510 & --- & --- & IRAC \\
{[3.6]} & 3.6 & post-gap & 10.4559 & 0.0049 & 0.0021 & IRAC \\
{[4.5]} & 4.5 & pre-gap & 10.4334 & --- & --- & IRAC \\
{[4.5]} & 4.5 & post-gap & 10.4367 & 0.0033 & 0.0021 & IRAC \\
\enddata
\tablenotetext{a}{Average magnitude.}
\tablenotetext{b}{Combined rms errors of the mean, i.e. rms scatter divided by the square root of (n-1) where n is the number of measurements.}
\end{deluxetable*}

\section{Discussion}

\subsection{Mass Limit of Circumstellar Dust}

Many of the hypotheses to explain the variability of KIC 8462852 depend on the presence of a substantial amount of circumstellar material. Excess emission from circumstellar dust is therefore an interesting diagnostic. Such an excess has not been found at a significant level \citep{lisse2015,marengo2015,boyajian2016,thompson2016}. However, previous searches have used data from 2MASS \citep{skrutskie2006}, {\it GALEX} \citep{morrissey2007}, warm {\it Spitzer} \citep{fazio2004}, {\it WISE} \citep{wright2010}, and new optical observations, which were taken more than 10 years apart, to constrain the stellar atmospheric models that are used to look for excess. The variability of the star, particularly if it has been fading for years \citep{montet2016}, could undermine the searches for excesses. 

Now we can test if KIC 8462852 has had a significant excess at 3.6 and 4.5 $\micron$ by fitting stellar models with the {\it Swift} data taken at specific epochs and compare the model-predicted stellar IR flux with the simultaneous {\it Spitzer} measurements. By October 2016, there have been five epochs at which we have {\it Swift} and {\it Spitzer} observations taken within 24 h: MJD 57403, 57454, 57621, 57635, and 57673. The first two are before March 2016, whereas the last three epochs are observed after the gap. To analyze these results, we adopt the ATLAS9 model \citep{castelli2004} with the stellar parameters obtained from spectroscopic observations, $\log (g) = 4.0$, and [M/H] = 0.0. We allow $T_{eff}$ to vary between 6700 and 7300 K with an increment of 100 K. We find that the stellar models with $T_{eff} = 7000$ and $7100$ K provide the best fits to the {\it Swift} photometry at all five epochs, with a minimum reduced $\chi^2$ from 0.08 to 1.5 (see Figure 2). We did not use the {\it Spitzer}/IRAC measurements to further constrain the fit, since we did not want to bias any evidence for an infrared excess. Although these temperatures are slightly higher than originally estimated by \citet{boyajian2016}, they are in reasonable agreement with the SED model based on the 2MASS photometry \citep[$T_{eff} = 6950$ K,][]{marengo2015} and the recent IRTF/SpeX spectrum leading to a classification of F1V - F2V, i.e., $\sim$ 6970 K  \citep{lisse2015}.  All of these values can be somewhat degenerate with changes in the assumed  logg and metallicity. For our purposes, however, having a good empirical fit into the ultraviolet allows placing constraints on the extinction. All five best-fit stellar models, one for each epoch, have $A_V$ in the range of 0.68 to 0.78, 2.0 to 2.3 times higher than $A_V = 0.341$ found by \citet{boyajian2016} with photometric measurements years apart.
In the {\it Spitzer}/IRAC wavebands at 3.6 and 4.5 $\micron$, the observed flux densities match the model-predicted stellar output fairly well at all five epochs. The average excess is $-0.39 \pm 0.30$ mJy at 3.6 $\micron$ and $-0.29 \pm 0.21$ mJy at 4.5 $\micron$. We conclude that we do not detect any significant excess of KIC 8462852 with near-simultaneous {\it Swift} and {\it Spitzer} observations.  Our conclusion is consistent with that of \citet{boyajian2016}, showing that it is independent of the uncertainties in fitting the stellar SED.

\begin{figure}[!t]
\epsscale{1.1}
\plotone{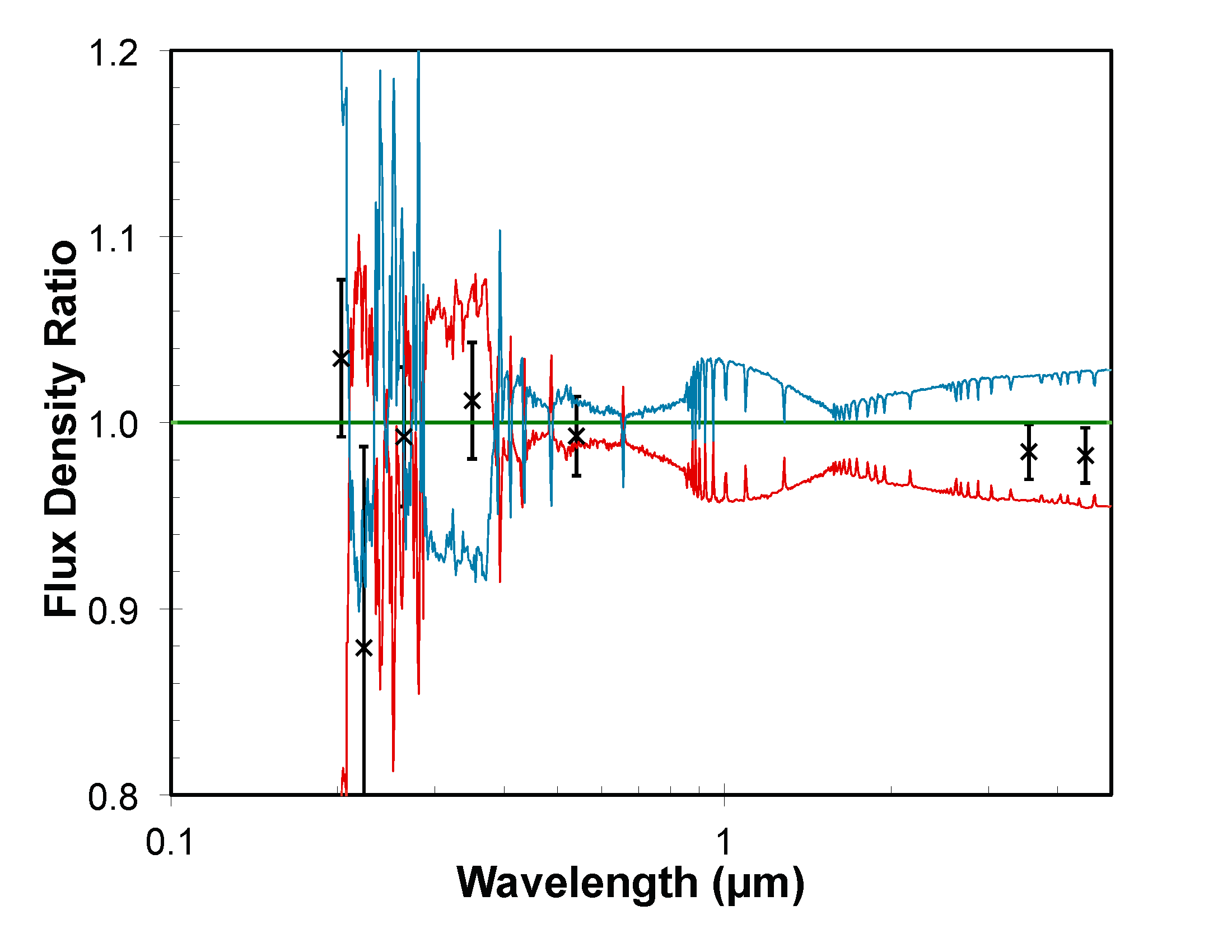}
\caption{Residuals to fits to stellar SEDs. The theoretical SEDs have been divided by that for 7000 K, which defines the best fit; this result is indicated by the horizontal green line. The cyan line is for $T_{eff}$ = 7300 K and the red one for $T_{eff}$ = 6750 K as adopted by \citet{boyajian2016}. The points with error bars show the UVOT and IRAC data. The BVR photometry is not shown because it is only the data at the short (UV) and long (IR) wavelength ends of the fit that have leverage on the assigned $T_{eff}$. The fits were carried out just on the UVOT photometry; the error bars on the IRAC photometry in the infrared include the expected uncertainties in the fit. 
\label{mass}}
\end{figure}

We have tested these conclusions using the average post-gap B, V, R, [3.6], and [4.5] measurements (Table 7) rather than the UVOT UV ones, standard stellar colors \citep{mamajek2017}, and a standard extinction law \citep{rieke1985,chapman2009}. The best fit was obtained assuming the star is of F1V spectral type (nominal temperature of 7030 K), with $A_V$ = 0.61. Given the uncertainties in the extinction law, the stellar models, and intrinsic stellar colors (e.g., the effects of metallicity), this agreement is excellent. The assigned extinction level also agrees roughly with the relatively red color of the star relative to a F2V comparison star in infrared spectra (C. M. Lisse, private communication). The conclusion about the absence of any infrared excess is unmodified with this calculation.

Upper limits to the level of circumstellar dust were also determined at 850 $\mu$m with JCMT/SCUBA-2 \citep{thompson2016}, and from WISE at 12 and 22 $\mu$m, in all cases where the stellar variations are relatively unimportant. Assuming that the grains emit as blockbodies and are distributed in a narrow, optically thin ring at various radii from the star, we place upper limits of $\sim$  4 $\times$ 10$^{-4}$ for the fractional luminosity, $L_{dust}/L_{*}$, for warm rings of radii between 0.1 and 10 AU and an order of magnitude higher for cold rings lying between 40 and 100 AU, which would be a typical cold-ring size for a star of this luminosity (see also \citet{boyajian2016}). These limits are consistent with the presence of a prominent debris disk, since even around young stars these systems usually have $L_{dust}/L_* \lesssim 10^{-3}$ \citep{wyatt2007,kenyon2008}. 

We carried out a second calculation to place upper limits on the possible dust masses. We again assumed that the dust is distributed in an optically thin ring (0.1 AU wide in the ring plane). We took the optical constants derived for debris disk material \citep{ballering2016} and used the Debris Disk Radiative Transfer Simulator\footnote{\url{http://www1.astrophysik.uni-kiel.de/dds/}} \citep{wolf2005}. The minimum grain radius was set to 1 $\micron$, roughly the blowout size for spherical particles around a F1/2 V star, and the power law index of the particle distribution was taken as 3.65 \citep{gaspar2012}. We take the upper limit at 850 $\micron$ to be 4.76 mJy, or 5.6$\sigma$ above zero, i.e., 3$\sigma$ added to the 2.6$\sigma$ ``signal'' at the position of the star \citep[rather than the 3$\sigma$ above zero as in][]{thompson2016}. Similarly, we take a 3$\sigma$ upper limit of 0.63 mJy at 4.5 $\micron$. We took the cataloged upper limits for the two WISE bands, which are computed in a similar way but at a 2$\sigma$ level. Stellar parameters are assumed to be $T_{eff} = 7000$ K and $L_* = 5$ L$_{\sun}$ \citep{pecaut2013,mamajek2017}.

\begin{figure*}[!t]
\epsscale{0.75}
\plotone{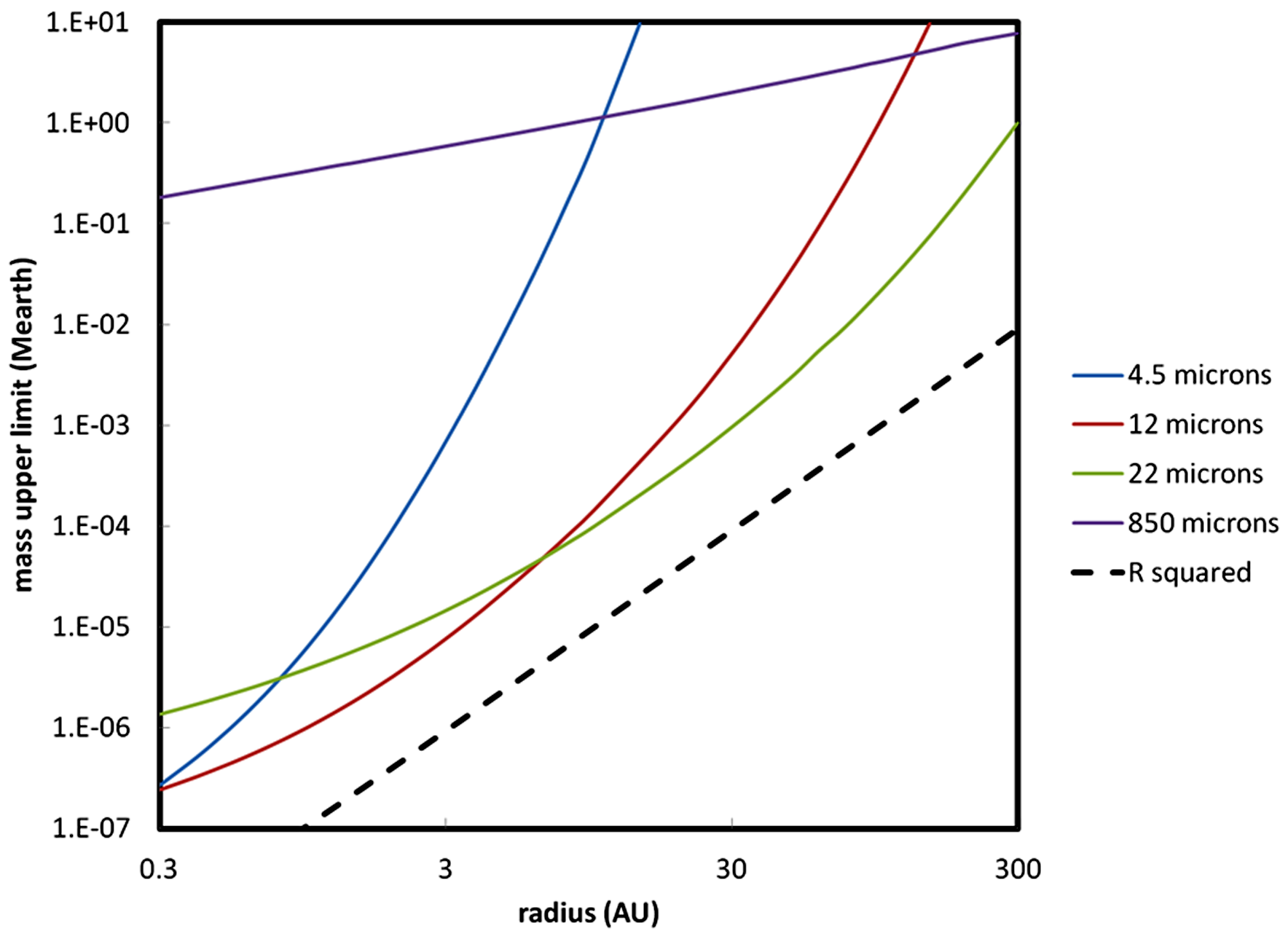}
\caption{Mass constraints on the circumstellar dust of KIC 8462852, computed by integrating the dust size distribution from the blowout size of 1 $\micron$ up to 2 mm. The dashed line goes as radius squared, for comparison with the similar behavior of the mass limits.
\label{mass}}
\end{figure*}

The resulting limits are shown in Figure~\ref{mass}. The upper limit at 100 AU is slightly higher than the mass of 0.017 M$_{\earth}$ for a similar range of dust sizes (i.e., up to 2 mm) in the Fomalhaut debris ring \citep{boley2012}. Again, a prominent but not extraordinary debris disk is allowed, corresponding for example to 2 - 3 orders of magnitude more dust than orbits the sun. The upper limits also permit sufficient dust mass to yield significant extinction. To demonstrate, we 
assume that the dust is in a ring at radius R with a thickness perpendicular to the orbital plane of 0.1 R; for simplicity we take an ISM-like dust particle size distribution with constant density within this ring. The resulting upper limit on the extinction is $A_V \sim 0.1$  \citep{guver2009}. This value is independent of the radius assumed for the ring, since as shown in Figure 3, the upper limits for the mass scale roughly as $R^2$, which is also the scaling of the ring area under our assumptions. Of course, this is only a rough estimate, but it is sufficient to demonstrate that detectable levels of extinction can be consistent with the upper limits on the thermal emission of any material surrounding the star.  That is, current measurements allow enough material to orbit KIC 8462852 to account for a number of the hypotheses for its behavior, such as the inspiralling and disintegration of massive comets. The minimum mass required to account for the long-term secular dimming 
through extinction is also within these mass constraints.  

\subsection{Extinction Curve}

We now explore the hypothesis that the long-term secular dimming of KIC 8462852 is due to variable extinction by dust in the line of sight. The absence of excess emission at 3.6 and 4.5 $\micron$ means that the photometry at these wavelengths is a measure of the stellar photospheric emission. Under the assumption that the fading of the star indicated in Table~\ref{color} is due to dust passing in front of the star, the relative amounts of dimming at the different wavelengths can therefore be used to constrain the wavelength dependence of the extinction from the UV to 4.5 $\mu$m in the IR. 

Under this hypothesis, the dimming of KIC 8462852 may arise either from the interstellar medium (ISM) or circumstellar material. For convenience, we describe the color of the fading in the terminology of interstellar extinction, although circumstellar material might have different behavior if the color were measured to high accuracy.   The Galactic ISM extinction curve from 0.1 to 3 $\micron$ can be well characterized by only one free parameter, the total-to-selective extinction ratio, defined as $R_V = A_V / E(B - V)$ \citep{cardelli1989}. Longward of 3 $\micron$, measurements towards diffuse ISM in the Galactic plane \citep{indebetouw2005}, towards the Galactic center \citep{fritz2011}, and towards dense molecular clouds in nearby star-forming regions \citep{chapman2009} reveal consistent shallow wavelength dependence of the ISM extinction in the {\it Spitzer}/IRAC bands. We find that extrapolating the analytical formula in \citet{cardelli1989} (CCM89,\defcitealias{cardelli1989}{CCM89}hereafter) to the 3.6 and 4.5 $\micron$ bands of IRAC yields $A_{[3.6]}/A_{K_s}$ from 0.42 to 0.54 and $A_{[4.5]}/A_{K_s}$ from 0.28 to 0.37 for $R_V$ values from 2.5 to 5.0, a range suitable for most sight lines in the Milky Way. Although the \citetalias{cardelli1989} extinction law does not claim to apply to these wavelengths, the 3.6 and 4.5 $\micron$ extrapolations are in good agreement with the IRAC observations \citep[cf. Table 3 in][]{chapman2009}. Therefore, for simplicity we adopt the \citetalias{cardelli1989} extinction law for all the seven bands monitored.

We have fitted the wavelength-dependent dimming in Table~\ref{color} with extinction curves using the formulation in \citetalias{cardelli1989},  parameterized by $R_V$ and extended to 4.5 $\mu$m. Figure 4 shows the results. Because of the relatively small level of dimming in the ultraviolet, the best-fitting extinction curves are relatively `gray', i.e., have large values of $R_V$. The vertical dashed lines show confidence levels corresponding to 1, 2, and 3 $\sigma$. Values of $R_V \sim 5$ are favored; the general value for the ISM, $R_V = 3.1$ is disfavored at a confidence level $>$ 90\%. Extinction even more gray than given by $R_V = 5$ (or a form differing more fundamentally from the interstellar law) is a definite possibility. However, a completely neutral extinction law is excluded because of the small variations at [3.6] and [4.5].

\begin{figure}[!t]
\epsscale{1.1}
\plotone{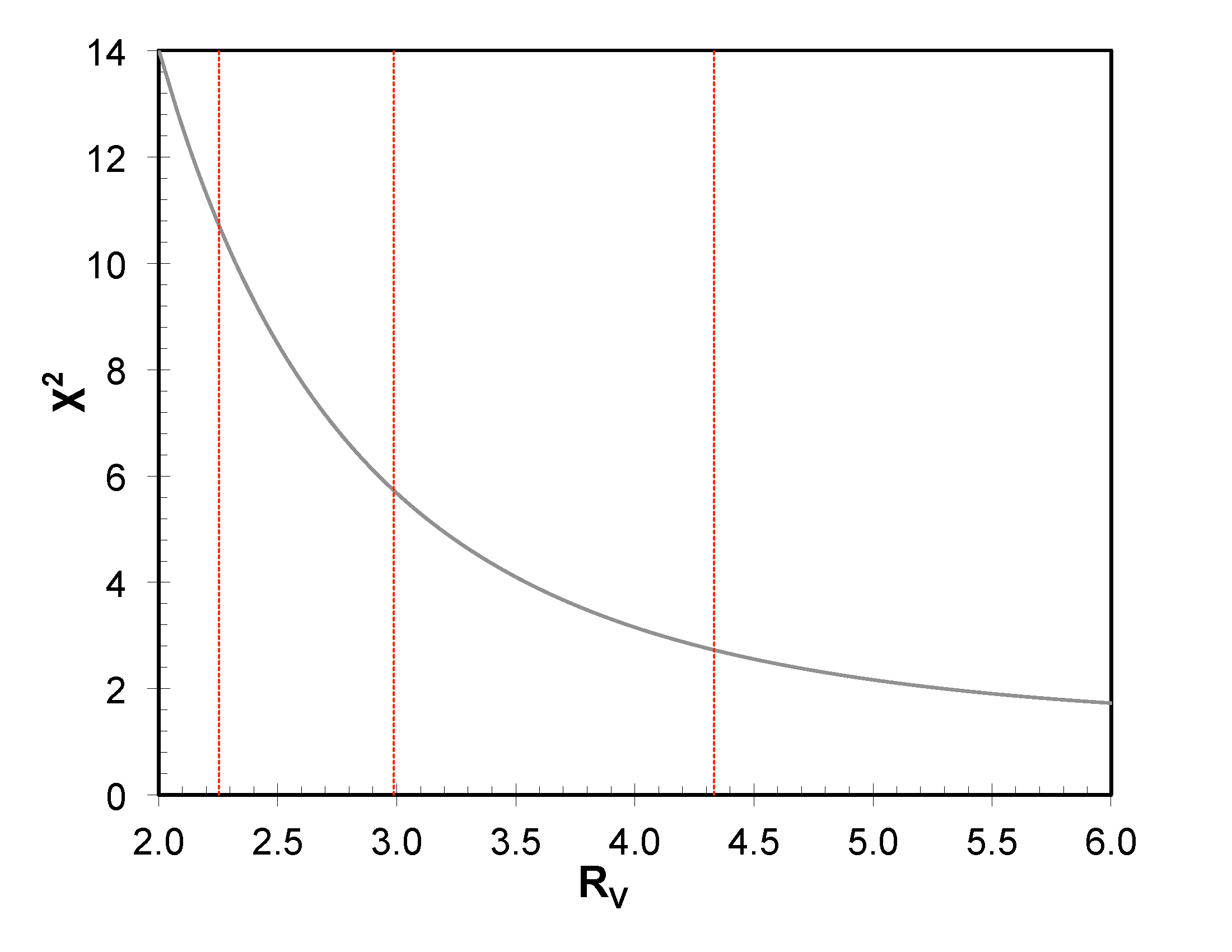}
\caption{Quality of the fits to the colors of the dimming of KIC 8462852 summarized in Table ~\ref{color}, as a function of the assumed value of $R_V$. The vertical lines are (from the right) at confidence levels corresponding respectively 
to 1, 2, and 3 $\sigma$. 
\label{RV}}
\end{figure}


\begin{figure*}[!t]
\epsscale{0.6}
\plotone{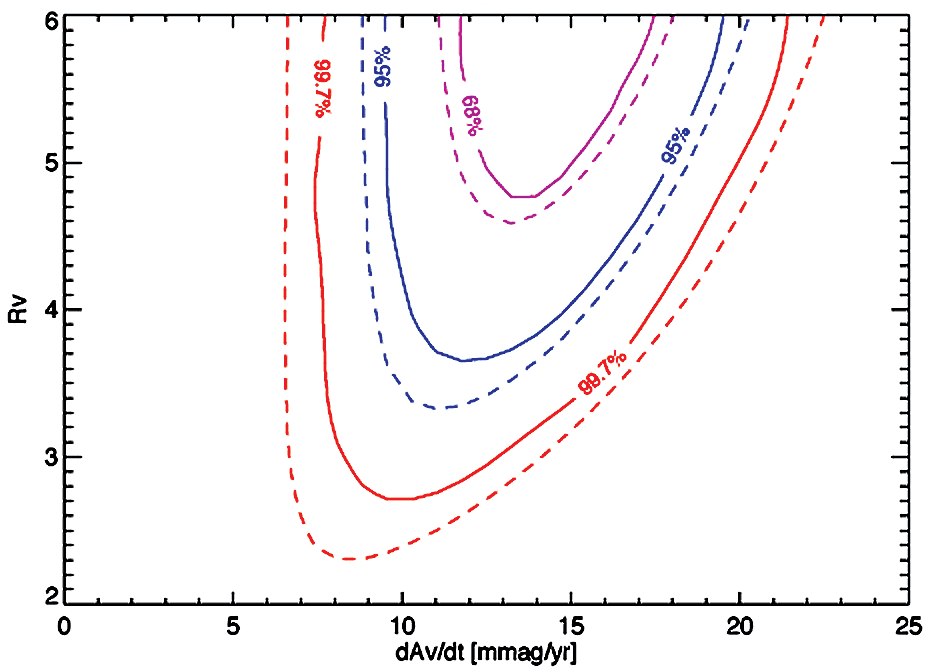}
\caption{Feldman and Cousins (F\&C) analysis of the average dimming rate (in units of $dA_V/dt$) and total-to-selective dimming ratio $R_V$. The graph shows the confidence levels (1, 2 and 3 $\sigma$) for the two fitting parameters derived from all data (all Swift/UVOT, High-Quality AstroLAB IRIS $BVR$, all IRAC except the SpiKeS epoch), derived by $\chi^2$ minimization (dashed lines) and through the F\&C method (solid lines). The F\&C method results in tighter confidence levels than do typical $\chi^2$ analyses, and provides a more reliable lower limit for $R_V$.}

\end{figure*}

So far we have conducted simple fits via $\chi^2$ minimization to the fading and extinction curves individually. However, they are interrelated. Therefore, we now fit them simultaneously using the Feldman and Cousins method (F\&C, Feldman \& Cousins, 1998; Sanchez et al. 2003). The method differs from a regular $\chi^2$ minimization by allowing setting of physical boundaries in the fitting parameters (e.g. $2 < R_V < 6$), and by adjusting the $\chi^2$ statistics accordingly, via a Monte Carlo approach. Figure 5 shows the resulting confidence intervals. Those for $\chi^2$ minimization agree excellently with the simple single-parameter results in Figure 4. The F\&C formalism suggests a similar conclusion, i.e. $R_V > 3.1$, but at somewhat higher confidence, $> 98$\%. The derived best fit dimming rate also agrees with the simple $\chi^2$ analysis. We also tried to fit the pre- and post-gap data separately. The result was not successful: the F\&C method fails to constrain reasonable values of $R_V$ and $dA_V /dt$. This suggests that there is no measurable dimming in the pre- and post-gap datasets separately (e.g. the dimming happened during the gap), which is also in agreement with the findings from simple $\chi^2$ minimization.

\section{Conclusions}

This paper continues the study of the long-term secular dimming of KIC 8462852, such as that seen during the Kepler mission \citep{montet2016}. 
We have observed a second dimming occurence, similar to that seen with Kepler. Our data extend from the UV (0.20 $\mu$m) to the mid-infrared (4.5 $\mu$m), allowing us to determine the spectral character of this event.  The dimming is less in the infrared than in the visible and UV, showing that the responsible bodies must be small, no more than a few microns in size. We analyze the colors under the assumption that the dimming is due to extinction by intervening dust. We find that the colors are likely to be more neutral than the reddening by typical sight lines in the ISM (confidence level $>$ 90\%). That is, the dust responsible for the dimming differs from that along typical sight lines in the ISM, suggesting that the dust is not of normal interstellar origin. 

The discovery of a dimming pattern similar to that seen with {\it Kepler} roughly 1400 days previously \citep{montet2016}  is challenging to reconcile with the hypothesis that these events result from dust produced during the assimilation of a planet \citep{wright2016,metzger2017}. 
The long-term secular dimming could correspond to some dusty structure in the Oort Cloud of the Sun with a column density gradient on $\sim$1 AU scale. The high ecliptic latitude of KIC 8462852 ($\beta = +62.2\degr$) is not necessarily a problem for this hypothsis, as the Oort cloud should be nearly isotropic \citep[][and references therein]{dones2015}. The primary difficulty with this picture is that the orbital timescale of any Oort cloud dust concentrations is $10^5$ to $10^7$ yr. Over the 8-year-long time line from the beginning of {\it Kepler} to our latest observations, the astrometric movement of such a structure should be dominated by the Earth's parallactic motion, and thus most of the observed light curve features should be recurrent relatively accurately on a yearly basis \citep{wright2016}. 

We conclude that  extinction by some form of circumstellar material is the most likely explanation for the long-term secular dimming.

\section*{Acknowledgements}
This work is based in part on observations made with the Spitzer Space Telescope, which is operated by the Jet Propulsion Laboratory, California Institute of Technology, under a contract with NASA. Support for this work was provided by NASA through an award issued by JPL/Caltech. GMK is supported by the Royal Society as a Royal Society University Research Fellow.
We acknowledge with thanks the variable star observations from the AAVSO International Database and the infrastructure maintained by AAVSO, which were used in this research. We thank Professor Jason Wright for his contiributions in acquiring the data. This publication makes use of data products from the Two Micron All Sky Survey, which is a joint project of the University of Massachusetts and the Infrared Processing and Analysis Center/California Institute of Technology, funded by the National Aeronautics and Space Administration and the National Science Foundation. This publication also makes use of data products from the Wide-field Infrared Survey Explorer, which is a joint project of the University of California, Los Angeles, and the Jet Propulsion Laboratory/California Institute of Technology, funded by the National Aeronautics and Space Administration.

\facility{AAVSO, Spitzer (IRAC), Swift (UVOT, XRT)}

\section{Appendix}

The following table provides the full set of UVOT photometry:

\begin{deluxetable*}{cccccccc}
\tabletypesize{\scriptsize}
\tablecaption{UVOT Photometry of KIC 8462852
\label{photom2}}
\tablehead{
\colhead{Waveband}& \colhead{MJD\tablenotemark{a}} & \colhead{raw} & \colhead{$\sigma$} & \colhead{corrected}  & \colhead{$\sigma$} & \colhead{relative comparison\tablenotemark{b}} & \colhead{$\sigma$}  \\
\colhead{}& \colhead{} & \colhead{magnitude} & \colhead{magnitude} & \colhead{magnitude}  & \colhead{magnitude} & \colhead{magnitude} & \colhead{magnitude}  
}
\startdata
$uvw2$	&	57317.518	&	14.820	&	0.050	&	14.776	&	0.089	&	0.045	&	0.074	\\
$uvw2$	&	57317.521	&	14.800	&	0.030	&	14.793	&	0.055	&	0.007	&	0.046	\\
$uvw2$	&	57317.552	&	14.790	&	0.030	&	14.787	&	0.047	&	0.003	&	0.036	\\
$uvw2$	&	57317.554	&	14.810	&	0.030	&	14.784	&	0.043	&	0.026	&	0.030	\\
$uvw2$	&	57317.584	&	14.780	&	0.030	&	14.755	&	0.058	&	0.025	&	0.050	\\
$uvw2$	&	57317.587	&	14.820	&	0.030	&	14.735	&	0.061	&	0.085	&	0.053	\\
$uvw2$	&	57357.014	&	14.790	&	0.030	&	14.772	&	0.049	&	0.018	&	0.039	\\
$uvw2$	&	57357.017	&	14.830	&	0.040	&	14.828	&	0.061	&	0.003	&	0.046	\\
$uvw2$	&	57360.465	&	14.790	&	0.040	&	14.802	&	0.063	&	-0.012	&	0.049	\\
$uvw2$	&	57363.857	&	14.870	&	0.040	&	14.802	&	0.067	&	0.068	&	0.054	\\
$uvw2$	&	57366.258	&	14.810	&	0.040	&	14.778	&	0.065	&	0.032	&	0.051	\\
$uvw2$	&	57369.981	&	14.770	&	0.040	&	14.791	&	0.061	&	-0.021	&	0.046	\\
$uvw2$	&	57369.983	&	14.760	&	0.040	&	14.746	&	0.064	&	0.014	&	0.050	\\
$uvw2$	&	57372.245	&	14.780	&	0.040	&	14.790	&	0.066	&	-0.010	&	0.053	\\
$uvw2$	&	57375.302	&	14.760	&	0.040	&	14.750	&	0.064	&	0.010	&	0.050	\\
$uvw2$	&	57378.828	&	14.810	&	0.040	&	14.799	&	0.063	&	0.011	&	0.049	\\
$uvw2$	&	57381.281	&	14.790	&	0.030	&	14.843	&	0.048	&	-0.053	&	0.037	\\
$uvw2$	&	57381.284	&	14.880	&	0.040	&	14.865	&	0.063	&	0.015	&	0.049	\\
$uvw2$	&	57384.147	&	14.760	&	0.040	&	14.791	&	0.063	&	-0.031	&	0.049	\\
$uvw2$	&	57387.934	&	14.870	&	0.040	&	14.951	&	0.060	&	-0.081	&	0.045	\\
$uvw2$	&	57390.524	&	14.910	&	0.040	&	14.979	&	0.063	&	-0.069	&	0.049	\\
$uvw2$	&	57393.641	&	14.760	&	0.030	&	14.766	&	0.045	&	-0.006	&	0.034	\\
$uvw2$	&	57393.645	&	14.810	&	0.040	&	14.870	&	0.064	&	-0.060	&	0.049	\\
$uvw2$	&	57396.567	&	14.800	&	0.040	&	14.876	&	0.063	&	-0.076	&	0.049	\\
$uvw2$	&	57399.031	&	14.790	&	0.040	&	14.736	&	0.065	&	0.054	&	0.051	\\
$uvw2$	&	57402.617	&	14.830	&	0.040	&	14.865	&	0.062	&	-0.035	&	0.047	\\
$uvw2$	&	57405.339	&	14.790	&	0.030	&	14.907	&	0.046	&	-0.117	&	0.035	\\
$uvw2$	&	57405.342	&	14.840	&	0.040	&	14.796	&	0.064	&	0.044	&	0.050	\\
$uvw2$	&	57408.274	&	14.820	&	0.040	&	14.809	&	0.064	&	0.011	&	0.050	\\
$uvw2$	&	57411.458	&	14.830	&	0.040	&	14.857	&	0.063	&	-0.027	&	0.049	\\
$uvw2$	&	57414.395	&	14.790	&	0.050	&	14.793	&	0.091	&	-0.003	&	0.076	\\
$uvw2$	&	57416.976	&	14.820	&	0.040	&	14.856	&	0.065	&	-0.036	&	0.051	\\
$uvw2$	&	57420.169	&	14.930	&	0.040	&	14.924	&	0.064	&	0.006	&	0.050	\\
$uvw2$	&	57423.164	&	14.870	&	0.040	&	14.797	&	0.066	&	0.073	&	0.052	\\
$uvw2$	&	57426.828	&	14.810	&	0.040	&	14.746	&	0.064	&	0.064	&	0.050	\\
$uvw2$	&	57429.809	&	14.780	&	0.040	&	14.761	&	0.059	&	0.019	&	0.043	\\
$uvw2$	&	57429.811	&	14.750	&	0.040	&	14.761	&	0.062	&	-0.011	&	0.047	\\
$uvw2$	&	57432.017	&	14.820	&	0.040	&	14.811	&	0.066	&	0.010	&	0.053	\\
$uvw2$	&	57435.009	&	14.780	&	0.040	&	14.783	&	0.064	&	-0.003	&	0.050	\\
$uvw2$	&	57438.533	&	14.810	&	0.040	&	14.821	&	0.064	&	-0.011	&	0.050	\\
$uvw2$	&	57441.518	&	14.780	&	0.040	&	14.777	&	0.060	&	0.004	&	0.045	\\
$uvw2$	&	57441.520	&	14.820	&	0.040	&	14.790	&	0.065	&	0.030	&	0.052	\\
$uvw2$	&	57444.565	&	14.870	&	0.040	&	14.819	&	0.066	&	0.051	&	0.053	\\
$uvw2$	&	57447.422	&	14.830	&	0.040	&	14.841	&	0.065	&	-0.011	&	0.052	\\
$uvw2$	&	57450.279	&	14.800	&	0.040	&	14.741	&	0.064	&	0.060	&	0.050	\\
$uvw2$	&	57453.599	&	14.820	&	0.040	&	14.898	&	0.061	&	-0.078	&	0.046	\\
$uvw2$	&	57453.601	&	14.800	&	0.040	&	14.767	&	0.064	&	0.033	&	0.050	\\
$uvw2$	&	57459.388	&	14.820	&	0.040	&	14.892	&	0.064	&	-0.072	&	0.050	\\
$uvw2$	&	57462.313	&	14.840	&	0.040	&	14.844	&	0.063	&	-0.004	&	0.049	\\
$uvw2$	&	57465.238	&	14.760	&	0.040	&	14.774	&	0.061	&	-0.014	&	0.046	\\
$uvw2$	&	57465.240	&	14.800	&	0.040	&	14.823	&	0.063	&	-0.023	&	0.049	\\
$uvw2$	&	57468.099	&	14.780	&	0.040	&	14.802	&	0.064	&	-0.022	&	0.050	\\
$uvw2$	&	57471.164	&	14.790	&	0.060	&	14.788	&	0.104	&	0.003	&	0.085	\\
$uvw2$	&	57474.015	&	14.820	&	0.040	&	14.802	&	0.064	&	0.018	&	0.050	\\
$uvw2$	&	57621.142	&	14.790	&	0.040	&	14.813	&	0.061	&	-0.023	&	0.046	\\
$uvw2$	&	57621.144	&	14.870	&	0.040	&	14.972	&	0.061	&	-0.102	&	0.046	\\
$uvw2$	&	57635.448	&	14.770	&	0.040	&	14.839	&	0.060	&	-0.069	&	0.045	\\
$uvw2$	&	57656.697	&	14.740	&	0.040	&	14.700	&	0.066	&	0.040	&	0.052	\\
$uvw2$	&	57672.645	&	14.860	&	0.040	&	14.917	&	0.065	&	-0.057	&	0.051	\\
$uvw2$	&	57677.098	&	14.880	&	0.040	&	14.932	&	0.061	&	-0.052	&	0.046	\\
$uvw2$	&	57677.101	&	14.720	&	0.040	&	14.830	&	0.061	&	-0.110	&	0.046	\\
$uvw2$	&	57708.730	&	14.750	&	0.050	&	14.840	&	0.082	&	-0.090	&	0.065	\\
$uvw2$	&	57719.421	&	14.780	&	0.040	&	14.785	&	0.066	&	-0.005	&	0.052	\\
$uvw2$	&	57723.277	&	14.780	&	0.040	&	14.810	&	0.061	&	-0.030	&	0.046	\\
$uvw2$	&	57730.131	&	14.740	&	0.040	&	14.743	&	0.066	&	-0.003	&	0.053	\\
$uvw2$	&	57737.433	&	14.720	&	0.070	&	14.896	&	0.113	&	-0.176	&	0.089	\\
$uvw2$	&	57737.450	&	14.790	&	0.040	&	14.867	&	0.061	&	-0.077	&	0.046	\\
$uvw2$	&	57742.752	&	14.730	&	0.040	&	14.777	&	0.063	&	-0.047	&	0.049	\\
$uvw2$	&	57744.212	&	14.790	&	0.040	&	14.814	&	0.064	&	-0.024	&	0.050	\\
$uvw2$	&	57751.851	&	14.780	&	0.040	&	14.774	&	0.066	&	0.006	&	0.052	\\
$uvw2$	&	57756.312	&	14.790	&	0.040	&	14.855	&	0.064	&	-0.065	&	0.050	\\
$uvw2$	&	57765.268	&	14.750	&	0.040	&	14.796	&	0.063	&	-0.046	&	0.049	\\
$uvw2$	&	57765.271	&	14.760	&	0.030	&	14.767	&	0.049	&	-0.007	&	0.038	\\
$uvw2$	&	57765.274	&	14.850	&	0.040	&	14.831	&	0.066	&	0.020	&	0.053	\\
$uvw2$	&	57772.977	&	14.770	&	0.040	&	14.793	&	0.060	&	-0.023	&	0.045	\\
$uvw2$	&	57779.890	&	14.760	&	0.040	&	14.897	&	0.063	&	-0.137	&	0.049	\\
$uvw2$	&	57786.140	&	14.780	&	0.040	&	14.829	&	0.064	&	-0.049	&	0.050	\\
$uvw2$	&	57789.993	&	14.790	&	0.040	&	14.864	&	0.064	&	-0.074	&	0.050	\\
$uvw2$	&	57792.322	&	14.780	&	0.040	&	14.913	&	0.063	&	-0.133	&	0.048	\\
$uvw2$	&	57795.382	&	14.800	&	0.040	&	14.903	&	0.063	&	-0.103	&	0.049	\\
$uvw2$	&	57798.502	&	14.810	&	0.040	&	14.919	&	0.062	&	-0.109	&	0.047	\\
$uvw2$	&	57801.894	&	14.800	&	0.040	&	14.790	&	0.067	&	0.010	&	0.054	\\
$uvw2$	&	57804.938	&	14.730	&	0.040	&	14.735	&	0.065	&	-0.005	&	0.052	\\
$uvw2$	&	57807.728	&	14.760	&	0.030	&	14.827	&	0.049	&	-0.067	&	0.039	\\
$uvw2$	&	57810.385	&	14.830	&	0.040	&	14.848	&	0.066	&	-0.018	&	0.052	\\
$uvw2$	&	57853.037	&	14.740	&	0.040	&	14.825	&	0.061	&	-0.085	&	0.046	\\
$uvw2$	&	57853.039	&	14.720	&	0.030	&	14.750	&	0.047	&	-0.030	&	0.037	\\
$uvw2$	&	57853.043	&	14.760	&	0.040	&	14.798	&	0.063	&	-0.038	&	0.048	\\
$uvm2$	&	57317.528	&	14.730	&	0.040	&	14.819	&	0.111	&	-0.089	&	0.104	\\
$uvm2$	&	57357.022	&	14.790	&	0.040	&	14.923	&	0.120	&	-0.133	&	0.113	\\
$uvm2$	&	57360.470	&	14.660	&	0.040	&	14.678	&	0.141	&	-0.018	&	0.135	\\
$uvm2$	&	57363.862	&	14.850	&	0.050	&	15.042	&	0.133	&	-0.192	&	0.124	\\
$uvm2$	&	57366.256	&	14.710	&	0.040	&	14.697	&	0.129	&	0.013	&	0.122	\\
$uvm2$	&	57372.250	&	14.720	&	0.050	&	14.742	&	0.143	&	-0.022	&	0.134	\\
$uvm2$	&	57375.307	&	14.700	&	0.040	&	14.696	&	0.137	&	0.004	&	0.131	\\
$uvm2$	&	57378.825	&	14.730	&	0.040	&	14.985	&	0.090	&	-0.255	&	0.081	\\
$uvm2$	&	57378.833	&	14.740	&	0.040	&	14.715	&	0.141	&	0.025	&	0.135	\\
$uvm2$	&	57381.290	&	14.790	&	0.040	&	14.751	&	0.131	&	0.039	&	0.125	\\
$uvm2$	&	57384.152	&	14.770	&	0.050	&	14.916	&	0.128	&	-0.146	&	0.117	\\
$uvm2$	&	57387.939	&	14.860	&	0.050	&	14.818	&	0.136	&	0.042	&	0.127	\\
$uvm2$	&	57390.521	&	14.690	&	0.030	&	14.788	&	0.086	&	-0.098	&	0.081	\\
$uvm2$	&	57390.529	&	14.900	&	0.050	&	15.006	&	0.134	&	-0.106	&	0.124	\\
$uvm2$	&	57393.650	&	14.760	&	0.050	&	14.763	&	0.140	&	-0.003	&	0.131	\\
$uvm2$	&	57396.572	&	14.670	&	0.040	&	14.802	&	0.133	&	-0.132	&	0.127	\\
$uvm2$	&	57399.036	&	14.690	&	0.040	&	14.604	&	0.147	&	0.086	&	0.141	\\
$uvm2$	&	57402.614	&	14.740	&	0.040	&	14.765	&	0.101	&	-0.025	&	0.093	\\
$uvm2$	&	57402.622	&	14.740	&	0.040	&	14.866	&	0.123	&	-0.126	&	0.117	\\
$uvm2$	&	57405.347	&	14.850	&	0.050	&	15.068	&	0.120	&	-0.218	&	0.109	\\
$uvm2$	&	57416.981	&	14.770	&	0.050	&	14.785	&	0.140	&	-0.015	&	0.131	\\
$uvm2$	&	57426.826	&	14.830	&	0.060	&	14.562	&	0.200	&	0.268	&	0.191	\\
$uvm2$	&	57426.833	&	14.800	&	0.050	&	14.932	&	0.132	&	-0.132	&	0.122	\\
$uvm2$	&	57444.570	&	14.810	&	0.050	&	14.876	&	0.142	&	-0.066	&	0.133	\\
$uvm2$	&	57447.427	&	14.750	&	0.050	&	14.955	&	0.128	&	-0.205	&	0.117	\\
$uvm2$	&	57450.276	&	14.880	&	0.040	&	14.953	&	0.095	&	-0.073	&	0.087	\\
$uvm2$	&	57450.284	&	14.660	&	0.040	&	14.646	&	0.149	&	0.014	&	0.144	\\
$uvm2$	&	57453.606	&	14.700	&	0.050	&	14.925	&	0.134	&	-0.225	&	0.124	\\
$uvm2$	&	57459.393	&	14.840	&	0.050	&	15.158	&	0.122	&	-0.318	&	0.111	\\
$uvm2$	&	57462.310	&	14.770	&	0.040	&	14.902	&	0.112	&	-0.132	&	0.104	\\
$uvm2$	&	57465.246	&	14.790	&	0.050	&	14.835	&	0.134	&	-0.045	&	0.124	\\
$uvm2$	&	57474.013	&	14.680	&	0.050	&	14.765	&	0.139	&	-0.085	&	0.130	\\
$uvm2$	&	57474.021	&	14.820	&	0.050	&	14.721	&	0.155	&	0.099	&	0.147	\\
$uvm2$	&	57621.149	&	14.750	&	0.050	&	14.840	&	0.140	&	-0.090	&	0.131	\\
$uvm2$	&	57635.454	&	14.640	&	0.040	&	14.668	&	0.129	&	-0.028	&	0.123	\\
$uvm2$	&	57656.702	&	14.820	&	0.050	&	14.911	&	0.138	&	-0.091	&	0.128	\\
$uvm2$	&	57672.650	&	14.680	&	0.050	&	14.647	&	0.146	&	0.033	&	0.137	\\
$uvm2$	&	57677.106	&	14.720	&	0.040	&	14.785	&	0.131	&	-0.065	&	0.125	\\
$uvm2$	&	57708.734	&	14.620	&	0.060	&	14.766	&	0.176	&	-0.146	&	0.166	\\
$uvm2$	&	57719.426	&	14.690	&	0.040	&	14.938	&	0.122	&	-0.248	&	0.115	\\
$uvm2$	&	57730.124	&	14.760	&	0.040	&	14.933	&	0.102	&	-0.173	&	0.094	\\
$uvm2$	&	57730.128	&	14.740	&	0.040	&	14.731	&	0.107	&	0.009	&	0.099	\\
$uvm2$	&	57730.136	&	14.750	&	0.050	&	14.931	&	0.129	&	-0.181	&	0.119	\\
$uvm2$	&	57742.746	&	14.750	&	0.050	&	14.702	&	0.151	&	0.048	&	0.142	\\
$uvm2$	&	57742.749	&	14.710	&	0.040	&	14.844	&	0.100	&	-0.134	&	0.092	\\
$uvm2$	&	57742.757	&	14.720	&	0.050	&	14.740	&	0.139	&	-0.020	&	0.130	\\
$uvm2$	&	57772.983	&	14.740	&	0.040	&	14.904	&	0.115	&	-0.164	&	0.107	\\
$uvm2$	&	57786.137	&	14.740	&	0.040	&	14.878	&	0.097	&	-0.138	&	0.088	\\
$uvm2$	&	57789.986	&	14.800	&	0.040	&	14.846	&	0.100	&	-0.046	&	0.091	\\
$uvm2$	&	57789.990	&	14.740	&	0.040	&	14.880	&	0.099	&	-0.140	&	0.091	\\
$uvm2$	&	57792.315	&	14.750	&	0.040	&	14.782	&	0.102	&	-0.032	&	0.093	\\
$uvm2$	&	57792.319	&	14.770	&	0.040	&	14.866	&	0.098	&	-0.096	&	0.089	\\
$uvm2$	&	57795.376	&	14.760	&	0.040	&	14.811	&	0.110	&	-0.051	&	0.102	\\
$uvm2$	&	57795.379	&	14.800	&	0.040	&	14.908	&	0.097	&	-0.108	&	0.088	\\
$uvm2$	&	57798.495	&	14.710	&	0.040	&	14.809	&	0.106	&	-0.099	&	0.098	\\
$uvm2$	&	57798.498	&	14.800	&	0.040	&	14.794	&	0.104	&	0.006	&	0.096	\\
$uvm2$	&	57804.932	&	14.710	&	0.040	&	14.674	&	0.101	&	0.036	&	0.093	\\
$uvm2$	&	57804.935	&	14.720	&	0.040	&	14.874	&	0.099	&	-0.154	&	0.091	\\
$uvm2$	&	57810.378	&	14.700	&	0.040	&	14.864	&	0.097	&	-0.164	&	0.088	\\
$uvm2$	&	57810.382	&	14.730	&	0.040	&	14.844	&	0.098	&	-0.114	&	0.090	\\
$uvm2$	&	57819.013	&	14.680	&	0.040	&	14.886	&	0.093	&	-0.206	&	0.084	\\
$uvm2$	&	57819.017	&	14.820	&	0.040	&	14.935	&	0.097	&	-0.115	&	0.088	\\
$uvm2$	&	57853.048	&	14.750	&	0.040	&	14.697	&	0.140	&	0.054	&	0.134	\\
$uvw1$	&	57317.531	&	13.650	&	0.030	&	13.646	&	0.038	&	0.004	&	0.023	\\
$uvw1$	&	57357.025	&	13.630	&	0.030	&	13.635	&	0.038	&	-0.005	&	0.024	\\
$uvw1$	&	57360.473	&	13.660	&	0.030	&	13.644	&	0.040	&	0.016	&	0.026	\\
$uvw1$	&	57363.855	&	13.650	&	0.030	&	13.549	&	0.049	&	0.101	&	0.038	\\
$uvw1$	&	57363.864	&	13.660	&	0.030	&	13.655	&	0.042	&	0.005	&	0.029	\\
$uvw1$	&	57369.991	&	13.600	&	0.030	&	13.585	&	0.042	&	0.015	&	0.029	\\
$uvw1$	&	57372.253	&	13.650	&	0.030	&	13.695	&	0.040	&	-0.045	&	0.026	\\
$uvw1$	&	57375.299	&	13.620	&	0.030	&	13.635	&	0.037	&	-0.015	&	0.022	\\
$uvw1$	&	57375.310	&	13.600	&	0.030	&	13.588	&	0.040	&	0.012	&	0.026	\\
$uvw1$	&	57378.836	&	13.620	&	0.030	&	13.584	&	0.040	&	0.036	&	0.026	\\
$uvw1$	&	57381.292	&	13.660	&	0.030	&	13.629	&	0.040	&	0.031	&	0.026	\\
$uvw1$	&	57387.930	&	13.620	&	0.030	&	13.644	&	0.037	&	-0.024	&	0.022	\\
$uvw1$	&	57387.942	&	13.660	&	0.030	&	13.610	&	0.040	&	0.050	&	0.026	\\
$uvw1$	&	57390.532	&	13.720	&	0.030	&	13.754	&	0.040	&	-0.034	&	0.026	\\
$uvw1$	&	57393.652	&	13.610	&	0.030	&	13.631	&	0.040	&	-0.021	&	0.026	\\
$uvw1$	&	57396.575	&	13.630	&	0.030	&	13.673	&	0.040	&	-0.043	&	0.026	\\
$uvw1$	&	57399.028	&	13.620	&	0.020	&	13.619	&	0.028	&	0.001	&	0.020	\\
$uvw1$	&	57399.038	&	13.630	&	0.030	&	13.652	&	0.040	&	-0.022	&	0.026	\\
$uvw1$	&	57402.625	&	13.610	&	0.030	&	13.611	&	0.040	&	-0.001	&	0.026	\\
$uvw1$	&	57405.350	&	13.660	&	0.030	&	13.621	&	0.040	&	0.040	&	0.026	\\
$uvw1$	&	57408.282	&	13.720	&	0.030	&	13.700	&	0.042	&	0.020	&	0.029	\\
$uvw1$	&	57411.454	&	13.620	&	0.020	&	13.618	&	0.027	&	0.002	&	0.018	\\
$uvw1$	&	57411.465	&	13.700	&	0.030	&	13.684	&	0.042	&	0.016	&	0.029	\\
$uvw1$	&	57416.983	&	13.650	&	0.030	&	13.628	&	0.042	&	0.022	&	0.029	\\
$uvw1$	&	57420.176	&	13.680	&	0.030	&	13.667	&	0.042	&	0.013	&	0.029	\\
$uvw1$	&	57423.161	&	13.630	&	0.020	&	13.634	&	0.028	&	-0.004	&	0.020	\\
$uvw1$	&	57423.171	&	13.690	&	0.030	&	13.630	&	0.042	&	0.060	&	0.029	\\
$uvw1$	&	57426.836	&	13.680	&	0.030	&	13.716	&	0.040	&	-0.036	&	0.026	\\
$uvw1$	&	57429.819	&	13.610	&	0.030	&	13.628	&	0.040	&	-0.018	&	0.026	\\
$uvw1$	&	57432.024	&	13.660	&	0.030	&	13.658	&	0.042	&	0.002	&	0.029	\\
$uvw1$	&	57435.016	&	13.660	&	0.030	&	13.695	&	0.044	&	-0.035	&	0.032	\\
$uvw1$	&	57441.528	&	13.700	&	0.030	&	13.675	&	0.042	&	0.025	&	0.029	\\
$uvw1$	&	57444.572	&	13.660	&	0.030	&	13.679	&	0.040	&	-0.019	&	0.026	\\
$uvw1$	&	57447.419	&	13.640	&	0.030	&	13.657	&	0.036	&	-0.017	&	0.020	\\
$uvw1$	&	57447.430	&	13.620	&	0.030	&	13.580	&	0.042	&	0.040	&	0.029	\\
$uvw1$	&	57450.287	&	13.660	&	0.030	&	13.683	&	0.041	&	-0.023	&	0.028	\\
$uvw1$	&	57453.608	&	13.630	&	0.030	&	13.677	&	0.041	&	-0.047	&	0.028	\\
$uvw1$	&	57459.386	&	13.690	&	0.030	&	13.647	&	0.044	&	0.043	&	0.032	\\
$uvw1$	&	57459.395	&	13.670	&	0.030	&	13.722	&	0.040	&	-0.052	&	0.026	\\
$uvw1$	&	57462.320	&	13.640	&	0.030	&	13.668	&	0.040	&	-0.028	&	0.026	\\
$uvw1$	&	57465.248	&	13.630	&	0.030	&	13.613	&	0.040	&	0.017	&	0.026	\\
$uvw1$	&	57468.107	&	13.630	&	0.030	&	13.634	&	0.040	&	-0.004	&	0.026	\\
$uvw1$	&	57474.023	&	13.670	&	0.030	&	13.673	&	0.040	&	-0.003	&	0.026	\\
$uvw1$	&	57621.151	&	13.640	&	0.030	&	13.687	&	0.040	&	-0.047	&	0.026	\\
$uvw1$	&	57635.446	&	13.630	&	0.030	&	13.676	&	0.038	&	-0.046	&	0.024	\\
$uvw1$	&	57635.456	&	13.610	&	0.030	&	13.667	&	0.039	&	-0.057	&	0.025	\\
$uvw1$	&	57656.705	&	13.590	&	0.030	&	13.659	&	0.040	&	-0.069	&	0.026	\\
$uvw1$	&	57672.652	&	13.640	&	0.030	&	13.642	&	0.042	&	-0.002	&	0.029	\\
$uvw1$	&	57677.109	&	13.580	&	0.030	&	13.552	&	0.042	&	0.028	&	0.029	\\
$uvw1$	&	57708.725	&	13.630	&	0.030	&	13.642	&	0.046	&	-0.012	&	0.035	\\
$uvw1$	&	57719.418	&	13.600	&	0.030	&	13.594	&	0.037	&	0.006	&	0.022	\\
$uvw1$	&	57719.429	&	13.640	&	0.030	&	13.635	&	0.042	&	0.005	&	0.029	\\
$uvw1$	&	57723.274	&	13.590	&	0.030	&	13.613	&	0.038	&	-0.023	&	0.024	\\
$uvw1$	&	57723.285	&	13.640	&	0.030	&	13.663	&	0.038	&	-0.023	&	0.024	\\
$uvw1$	&	57730.139	&	13.640	&	0.030	&	13.691	&	0.040	&	-0.051	&	0.026	\\
$uvw1$	&	57742.760	&	13.620	&	0.030	&	13.627	&	0.040	&	-0.007	&	0.026	\\
$uvw1$	&	57744.219	&	13.610	&	0.030	&	13.656	&	0.040	&	-0.046	&	0.026	\\
$uvw1$	&	57751.845	&	13.590	&	0.030	&	13.654	&	0.039	&	-0.064	&	0.025	\\
$uvw1$	&	57751.848	&	13.660	&	0.030	&	13.672	&	0.037	&	-0.012	&	0.022	\\
$uvw1$	&	57751.859	&	13.640	&	0.030	&	13.677	&	0.041	&	-0.037	&	0.028	\\
$uvw1$	&	57756.319	&	13.640	&	0.030	&	13.680	&	0.042	&	-0.040	&	0.029	\\
$uvw1$	&	57772.986	&	13.620	&	0.030	&	13.651	&	0.039	&	-0.031	&	0.025	\\
$uvw1$	&	57779.884	&	13.600	&	0.030	&	13.642	&	0.038	&	-0.042	&	0.024	\\
$uvw1$	&	57779.887	&	13.630	&	0.030	&	13.668	&	0.037	&	-0.038	&	0.022	\\
$uvw1$	&	57779.898	&	13.600	&	0.030	&	13.635	&	0.042	&	-0.035	&	0.029	\\
$uvw1$	&	57786.148	&	13.620	&	0.030	&	13.672	&	0.042	&	-0.052	&	0.029	\\
$uvw1$	&	57790.001	&	13.620	&	0.030	&	13.655	&	0.042	&	-0.035	&	0.029	\\
$uvw1$	&	57792.329	&	13.640	&	0.030	&	13.693	&	0.042	&	-0.053	&	0.029	\\
$uvw1$	&	57795.389	&	13.620	&	0.030	&	13.650	&	0.042	&	-0.030	&	0.029	\\
$uvw1$	&	57798.510	&	13.600	&	0.030	&	13.672	&	0.040	&	-0.072	&	0.026	\\
$uvw1$	&	57804.946	&	13.620	&	0.030	&	13.636	&	0.042	&	-0.016	&	0.029	\\
$uvw1$	&	57807.739	&	13.640	&	0.030	&	13.677	&	0.038	&	-0.037	&	0.023	\\
$uvw1$	&	57810.392	&	13.620	&	0.030	&	13.659	&	0.040	&	-0.039	&	0.026	\\
$uvw1$	&	57819.027	&	13.580	&	0.030	&	13.645	&	0.041	&	-0.065	&	0.028	\\
$uvw1$	&	57853.051	&	13.610	&	0.030	&	13.628	&	0.040	&	-0.018	&	0.026	\\
$U$	&	57317.534	&	12.560	&	0.020	&	12.557	&	0.025	&	0.003	&	0.015	\\
$U$	&	57357.028	&	12.580	&	0.020	&	12.594	&	0.025	&	-0.014	&	0.015	\\
$U$	&	57360.463	&	12.560	&	0.030	&	12.563	&	0.035	&	-0.003	&	0.017	\\
$U$	&	57360.475	&	12.560	&	0.030	&	12.574	&	0.034	&	-0.014	&	0.016	\\
$U$	&	57363.867	&	12.580	&	0.030	&	12.567	&	0.034	&	0.013	&	0.016	\\
$U$	&	57369.993	&	12.560	&	0.030	&	12.559	&	0.034	&	0.001	&	0.015	\\
$U$	&	57372.243	&	12.570	&	0.030	&	12.584	&	0.034	&	-0.014	&	0.015	\\
$U$	&	57372.255	&	12.570	&	0.030	&	12.579	&	0.034	&	-0.009	&	0.015	\\
$U$	&	57375.313	&	12.570	&	0.020	&	12.574	&	0.025	&	-0.004	&	0.015	\\
$U$	&	57378.838	&	12.550	&	0.020	&	12.549	&	0.025	&	0.001	&	0.015	\\
$U$	&	57381.295	&	12.590	&	0.020	&	12.579	&	0.025	&	0.011	&	0.015	\\
$U$	&	57387.944	&	12.590	&	0.020	&	12.589	&	0.025	&	0.001	&	0.015	\\
$U$	&	57390.534	&	12.590	&	0.030	&	12.602	&	0.034	&	-0.012	&	0.015	\\
$U$	&	57393.654	&	12.580	&	0.030	&	12.564	&	0.034	&	0.016	&	0.015	\\
$U$	&	57396.565	&	12.590	&	0.030	&	12.574	&	0.034	&	0.016	&	0.015	\\
$U$	&	57396.577	&	12.580	&	0.030	&	12.594	&	0.034	&	-0.014	&	0.015	\\
$U$	&	57399.040	&	12.550	&	0.030	&	12.575	&	0.035	&	-0.025	&	0.017	\\
$U$	&	57402.627	&	12.570	&	0.020	&	12.589	&	0.025	&	-0.019	&	0.015	\\
$U$	&	57408.272	&	12.570	&	0.030	&	12.559	&	0.034	&	0.011	&	0.015	\\
$U$	&	57408.284	&	12.600	&	0.030	&	12.596	&	0.035	&	0.004	&	0.017	\\
$U$	&	57411.467	&	12.590	&	0.030	&	12.589	&	0.034	&	0.001	&	0.015	\\
$U$	&	57416.973	&	12.580	&	0.030	&	12.579	&	0.034	&	0.001	&	0.015	\\
$U$	&	57416.986	&	12.600	&	0.030	&	12.589	&	0.035	&	0.011	&	0.017	\\
$U$	&	57420.166	&	12.590	&	0.030	&	12.584	&	0.034	&	0.006	&	0.015	\\
$U$	&	57420.179	&	12.590	&	0.030	&	12.552	&	0.035	&	0.038	&	0.017	\\
$U$	&	57423.174	&	12.600	&	0.030	&	12.580	&	0.035	&	0.020	&	0.017	\\
$U$	&	57426.839	&	12.570	&	0.020	&	12.559	&	0.025	&	0.011	&	0.015	\\
$U$	&	57429.822	&	12.580	&	0.020	&	12.572	&	0.025	&	0.008	&	0.015	\\
$U$	&	57432.014	&	12.570	&	0.030	&	12.574	&	0.034	&	-0.004	&	0.015	\\
$U$	&	57432.027	&	12.600	&	0.030	&	12.605	&	0.034	&	-0.005	&	0.016	\\
$U$	&	57441.530	&	12.640	&	0.030	&	12.600	&	0.035	&	0.040	&	0.017	\\
$U$	&	57444.561	&	12.590	&	0.020	&	12.599	&	0.025	&	-0.009	&	0.015	\\
$U$	&	57444.574	&	12.560	&	0.030	&	12.557	&	0.034	&	0.003	&	0.015	\\
$U$	&	57447.432	&	12.580	&	0.030	&	12.582	&	0.034	&	-0.002	&	0.015	\\
$U$	&	57450.289	&	12.580	&	0.030	&	12.569	&	0.034	&	0.011	&	0.015	\\
$U$	&	57453.611	&	12.560	&	0.030	&	12.564	&	0.034	&	-0.004	&	0.015	\\
$U$	&	57459.397	&	12.580	&	0.030	&	12.607	&	0.034	&	-0.027	&	0.016	\\
$U$	&	57462.322	&	12.590	&	0.030	&	12.599	&	0.034	&	-0.009	&	0.016	\\
$U$	&	57465.251	&	12.590	&	0.020	&	12.604	&	0.025	&	-0.014	&	0.015	\\
$U$	&	57468.096	&	12.630	&	0.030	&	12.632	&	0.034	&	-0.002	&	0.015	\\
$U$	&	57468.109	&	12.580	&	0.020	&	12.572	&	0.025	&	0.008	&	0.015	\\
$U$	&	57474.026	&	12.600	&	0.020	&	12.601	&	0.026	&	-0.001	&	0.016	\\
$U$	&	57621.154	&	12.580	&	0.030	&	12.587	&	0.034	&	-0.007	&	0.015	\\
$U$	&	57635.459	&	12.580	&	0.020	&	12.599	&	0.025	&	-0.019	&	0.015	\\
$U$	&	57656.695	&	12.600	&	0.030	&	12.604	&	0.034	&	-0.004	&	0.015	\\
$U$	&	57656.707	&	12.590	&	0.030	&	12.597	&	0.034	&	-0.007	&	0.015	\\
$U$	&	57672.642	&	12.650	&	0.030	&	12.642	&	0.034	&	0.008	&	0.015	\\
$U$	&	57672.654	&	12.580	&	0.030	&	12.587	&	0.034	&	-0.007	&	0.015	\\
$U$	&	57677.111	&	12.590	&	0.030	&	12.622	&	0.034	&	-0.032	&	0.015	\\
$U$	&	57708.727	&	12.560	&	0.030	&	12.587	&	0.035	&	-0.027	&	0.017	\\
$U$	&	57719.431	&	12.570	&	0.020	&	12.557	&	0.025	&	0.013	&	0.015	\\
$U$	&	57723.288	&	12.590	&	0.020	&	12.592	&	0.025	&	-0.002	&	0.015	\\
$U$	&	57730.141	&	12.590	&	0.030	&	12.582	&	0.034	&	0.008	&	0.015	\\
$U$	&	57742.763	&	12.580	&	0.020	&	12.597	&	0.025	&	-0.017	&	0.015	\\
$U$	&	57744.209	&	12.600	&	0.020	&	12.604	&	0.025	&	-0.004	&	0.015	\\
$U$	&	57744.222	&	12.600	&	0.030	&	12.594	&	0.034	&	0.006	&	0.016	\\
$U$	&	57751.861	&	12.580	&	0.020	&	12.574	&	0.025	&	0.006	&	0.015	\\
$U$	&	57756.307	&	12.620	&	0.030	&	12.607	&	0.034	&	0.013	&	0.015	\\
$U$	&	57756.309	&	12.640	&	0.030	&	12.629	&	0.034	&	0.011	&	0.015	\\
$U$	&	57756.322	&	12.590	&	0.030	&	12.564	&	0.034	&	0.026	&	0.015	\\
$U$	&	57772.970	&	12.600	&	0.030	&	12.599	&	0.034	&	0.001	&	0.015	\\
$U$	&	57772.973	&	12.610	&	0.020	&	12.607	&	0.025	&	0.003	&	0.015	\\
$U$	&	57772.989	&	12.590	&	0.020	&	12.589	&	0.025	&	0.001	&	0.015	\\
$U$	&	57779.900	&	12.590	&	0.030	&	12.568	&	0.035	&	0.022	&	0.017	\\
$U$	&	57786.150	&	12.600	&	0.030	&	12.587	&	0.035	&	0.013	&	0.017	\\
$U$	&	57790.003	&	12.600	&	0.030	&	12.566	&	0.034	&	0.034	&	0.016	\\
$U$	&	57792.331	&	12.590	&	0.030	&	12.529	&	0.035	&	0.061	&	0.017	\\
$U$	&	57795.391	&	12.600	&	0.030	&	12.587	&	0.035	&	0.013	&	0.017	\\
$U$	&	57798.512	&	12.600	&	0.020	&	12.602	&	0.025	&	-0.002	&	0.015	\\
$U$	&	57804.948	&	12.590	&	0.030	&	12.577	&	0.034	&	0.013	&	0.015	\\
$U$	&	57807.743	&	12.590	&	0.020	&	12.587	&	0.025	&	0.003	&	0.015	\\
$U$	&	57810.395	&	12.590	&	0.030	&	12.597	&	0.034	&	-0.007	&	0.015	\\
$U$	&	57819.029	&	12.580	&	0.030	&	12.577	&	0.034	&	0.003	&	0.016	\\
$U$	&	57853.053	&	12.580	&	0.020	&	12.564	&	0.025	&	0.016	&	0.015	\\
$V$	&	57317.524	&	11.890	&	0.020	&	11.880	&	0.025	&	0.010	&	0.014	\\
$V$	&	57317.557	&	11.900	&	0.020	&	11.899	&	0.023	&	0.001	&	0.012	\\
$V$	&	57317.591	&	11.900	&	0.020	&	11.910	&	0.025	&	-0.010	&	0.014	\\
$V$	&	57357.020	&	11.890	&	0.020	&	11.878	&	0.023	&	0.012	&	0.012	\\
$V$	&	57360.468	&	11.880	&	0.020	&	11.880	&	0.023	&	0.000	&	0.012	\\
$V$	&	57363.860	&	11.910	&	0.020	&	11.888	&	0.023	&	0.022	&	0.012	\\
$V$	&	57366.260	&	11.880	&	0.030	&	11.850	&	0.035	&	0.031	&	0.019	\\
$V$	&	57369.986	&	11.890	&	0.020	&	11.887	&	0.023	&	0.003	&	0.012	\\
$V$	&	57372.248	&	11.900	&	0.020	&	11.906	&	0.023	&	-0.006	&	0.012	\\
$V$	&	57375.305	&	11.890	&	0.020	&	11.877	&	0.023	&	0.013	&	0.012	\\
$V$	&	57378.831	&	11.910	&	0.020	&	11.912	&	0.023	&	-0.002	&	0.012	\\
$V$	&	57381.287	&	11.900	&	0.020	&	11.901	&	0.023	&	-0.001	&	0.012	\\
$V$	&	57384.149	&	11.890	&	0.020	&	11.908	&	0.023	&	-0.018	&	0.012	\\
$V$	&	57387.936	&	11.900	&	0.020	&	11.900	&	0.023	&	0.000	&	0.012	\\
$V$	&	57390.527	&	11.920	&	0.020	&	11.916	&	0.023	&	0.004	&	0.012	\\
$V$	&	57393.647	&	11.900	&	0.020	&	11.903	&	0.023	&	-0.003	&	0.012	\\
$V$	&	57396.570	&	11.900	&	0.020	&	11.900	&	0.023	&	0.000	&	0.012	\\
$V$	&	57399.033	&	11.880	&	0.020	&	11.899	&	0.023	&	-0.019	&	0.012	\\
$V$	&	57402.619	&	11.890	&	0.020	&	11.887	&	0.023	&	0.003	&	0.012	\\
$V$	&	57405.345	&	11.920	&	0.020	&	11.916	&	0.023	&	0.004	&	0.012	\\
$V$	&	57408.277	&	11.920	&	0.020	&	11.916	&	0.023	&	0.004	&	0.012	\\
$V$	&	57411.460	&	11.920	&	0.020	&	11.913	&	0.023	&	0.008	&	0.012	\\
$V$	&	57416.978	&	11.890	&	0.020	&	11.891	&	0.023	&	-0.001	&	0.012	\\
$V$	&	57420.171	&	11.920	&	0.020	&	11.900	&	0.023	&	0.020	&	0.012	\\
$V$	&	57423.166	&	11.920	&	0.020	&	11.911	&	0.023	&	0.009	&	0.012	\\
$V$	&	57426.831	&	11.900	&	0.020	&	11.899	&	0.023	&	0.001	&	0.012	\\
$V$	&	57429.814	&	11.890	&	0.020	&	11.883	&	0.023	&	0.007	&	0.012	\\
$V$	&	57432.019	&	11.910	&	0.020	&	11.903	&	0.023	&	0.008	&	0.012	\\
$V$	&	57435.012	&	11.900	&	0.020	&	11.909	&	0.023	&	-0.009	&	0.012	\\
$V$	&	57438.535	&	11.910	&	0.020	&	11.911	&	0.023	&	-0.001	&	0.012	\\
$V$	&	57441.523	&	11.930	&	0.020	&	11.907	&	0.023	&	0.023	&	0.012	\\
$V$	&	57444.567	&	11.920	&	0.020	&	11.911	&	0.023	&	0.010	&	0.012	\\
$V$	&	57447.425	&	11.900	&	0.020	&	11.903	&	0.023	&	-0.003	&	0.012	\\
$V$	&	57450.282	&	11.930	&	0.020	&	11.926	&	0.023	&	0.004	&	0.012	\\
$V$	&	57453.603	&	11.890	&	0.020	&	11.903	&	0.023	&	-0.013	&	0.012	\\
$V$	&	57459.390	&	11.890	&	0.020	&	11.901	&	0.023	&	-0.011	&	0.012	\\
$V$	&	57462.315	&	11.900	&	0.020	&	11.911	&	0.023	&	-0.011	&	0.012	\\
$V$	&	57465.243	&	11.900	&	0.020	&	11.915	&	0.023	&	-0.015	&	0.012	\\
$V$	&	57468.101	&	11.910	&	0.020	&	11.921	&	0.023	&	-0.011	&	0.012	\\
$V$	&	57474.018	&	11.900	&	0.020	&	11.902	&	0.023	&	-0.002	&	0.012	\\
$V$	&	57621.146	&	11.890	&	0.020	&	11.904	&	0.023	&	-0.014	&	0.012	\\
$V$	&	57635.451	&	11.880	&	0.020	&	11.899	&	0.023	&	-0.019	&	0.012	\\
$V$	&	57656.700	&	11.890	&	0.020	&	11.898	&	0.023	&	-0.008	&	0.012	\\
$V$	&	57672.647	&	11.870	&	0.020	&	11.875	&	0.023	&	-0.005	&	0.012	\\
$V$	&	57677.103	&	11.880	&	0.020	&	11.897	&	0.023	&	-0.017	&	0.012	\\
$V$	&	57708.732	&	11.940	&	0.030	&	11.928	&	0.034	&	0.012	&	0.016	\\
$V$	&	57719.424	&	11.940	&	0.020	&	11.917	&	0.023	&	0.023	&	0.012	\\
$V$	&	57723.280	&	11.920	&	0.020	&	11.887	&	0.023	&	0.033	&	0.012	\\
$V$	&	57730.133	&	11.930	&	0.020	&	11.910	&	0.023	&	0.020	&	0.012	\\
$V$	&	57742.755	&	11.940	&	0.020	&	11.914	&	0.023	&	0.026	&	0.012	\\
$V$	&	57744.214	&	11.940	&	0.020	&	11.909	&	0.023	&	0.031	&	0.012	\\
$V$	&	57751.853	&	11.930	&	0.020	&	11.909	&	0.023	&	0.021	&	0.012	\\
$V$	&	57756.314	&	11.950	&	0.020	&	11.930	&	0.023	&	0.020	&	0.012	\\
$V$	&	57765.275	&	11.920	&	0.030	&	11.895	&	0.036	&	0.025	&	0.020	\\
$V$	&	57772.980	&	11.930	&	0.020	&	11.903	&	0.023	&	0.027	&	0.012	\\
$V$	&	57779.893	&	11.930	&	0.020	&	11.890	&	0.023	&	0.040	&	0.012	\\
$V$	&	57786.143	&	11.950	&	0.020	&	11.930	&	0.023	&	0.020	&	0.012	\\
$V$	&	57789.996	&	11.930	&	0.020	&	11.885	&	0.023	&	0.045	&	0.012	\\
$V$	&	57792.324	&	11.940	&	0.020	&	11.914	&	0.023	&	0.026	&	0.012	\\
$V$	&	57795.384	&	11.950	&	0.020	&	11.877	&	0.023	&	0.073	&	0.012	\\
$V$	&	57798.504	&	11.930	&	0.020	&	11.897	&	0.023	&	0.033	&	0.012	\\
$V$	&	57801.896	&	11.940	&	0.020	&	11.920	&	0.023	&	0.020	&	0.012	\\
$V$	&	57804.941	&	11.910	&	0.020	&	11.865	&	0.023	&	0.045	&	0.012	\\
$V$	&	57807.732	&	11.930	&	0.020	&	11.899	&	0.023	&	0.031	&	0.012	\\
$V$	&	57810.387	&	11.930	&	0.020	&	11.908	&	0.023	&	0.022	&	0.012	\\
$V$	&	57819.022	&	11.930	&	0.020	&	11.913	&	0.023	&	0.017	&	0.012	\\
$V$	&	57853.045	&	11.920	&	0.020	&	11.885	&	0.023	&	0.035	&	0.012	\\
\enddata
\tablenotetext{a}{Modified Julian date}
\tablenotetext{b}{Average brightening of the comparison star measurements}
\end{deluxetable*}

\end{document}